\documentclass[%
reprint,
superscriptaddress,
amsmath,amssymb,
aps,
prx,
]{revtex4-2}

\usepackage{graphicx}
\usepackage{dcolumn}
\usepackage{braket}
\usepackage{bm}
\usepackage{color}
\usepackage{multirow}
\usepackage{siunitx}

\usepackage{comment}

\newcommand{\gnd}{{}^{1}\!S_{0}}
\newcommand{\clock}{{}^{3}\!P_{0}}
\newcommand{\ex}{{}^{3}\!P_{1}}
\newcommand{\mj}[1]{m_j\!=\!#1}

\newcommand{\cTR}{\cos\!\left(\frac{\theta_R}{2}\right)} 
\newcommand{\sTR}{\sin\!\left(\frac{\theta_R}{2}\right)} 

\newcommand{\sT}{\sin\!\left(\frac{\theta_1}{2}\right)} 
\newcommand{\cT}{\cos\!\left(\frac{\theta_1}{2}\right)} 

\usepackage{hyperref}    
\hypersetup{
    colorlinks=true,       
    linkcolor=blue,        
    citecolor=blue,        
    urlcolor=blue          
}

\begin{document}


\title{Microsecond-Scale Coherent Control of a Forbidden Clock Transition with Doppler-Free Multiphoton Excitations}

\author{Guglielmo Panelli}%
\thanks{These authors contributed equally to this work.}%
\affiliation{%
 Department of Physics, Stanford University, Stanford, California 94305, USA 
}%
\author{Erik J. Porter}
\thanks{These authors contributed equally to this work.}%
\affiliation{%
 Department of Electrical Engineering, Stanford University, Stanford, California 94305, USA 
}%
\author{V. Rose Knight}%
\affiliation{%
 Department of Applied Physics, Stanford University, Stanford, California 94305, USA  
}%

\author{Shaun C. Burd}%
\affiliation{%
 Department of Physics, Stanford University, Stanford, California 94305, USA 
}%
\affiliation{SLAC National Accelerator Laboratory, Menlo Park, CA 94025, USA}
\author{Mark Kasevich}
\email{kasevich@stanford.edu}
\affiliation{%
 Department of Physics, Stanford University, Stanford, California 94305, USA  
}%
\affiliation{%
 Department of Applied Physics, Stanford University, Stanford, California 94305, USA  
}%

\date{\today}

\begin{abstract}
We demonstrate two Doppler-free (DF) excitation schemes for coherent manipulation of the $\gnd-\clock$ clock transition in $^{88}$Sr that achieve microsecond-scale excitation times.
The first approach uses three-photon excitation with distinct phase-coherent spectral components to couple the ground and clock states while canceling the first-order Doppler shift. 
The second approach is a sequential protocol that combines a single-photon excitation with a two-photon Raman excitation, reducing coherent clock transition manipulation times to below a microsecond.
With both methods, we perform high-contrast Ramsey spectroscopy on thermal ensembles of $3\times10^{6}$ atoms in free space. We observe three orders of magnitude suppression of Doppler dephasing compared to single-photon excitation, relaxing the need for tight confinement or ultra-low temperatures. 
These broadly applicable techniques enable fast, coherent manipulation of narrow-line transitions, with 
implications for optical atomic clocks, matter-wave interferometers, quantum-enhanced metrology, and quantum information processing.
\end{abstract}

                            
\maketitle

The ability to coherently control optical clock transitions underpins many of the recent advances in atomic sensors, quantum computing, and precision metrology~\cite{hollberg_optical_2005, Ludlow2015, Safronova2018, Saffman2010}. 
Leveraging the long coherence times of these ultra-narrow transitions imposes demanding technical requirements and stringent limits on achievable excitation rates.
To date, use of these transitions has generally been limited to ultracold or Lamb-Dicke-confined systems, including neutral atoms trapped in optical lattices or tweezers~\cite{Takamoto2005, Saffman2010, Norcia2019}, Bose-Einstein condensates~\cite{Ido2005}, and ions in electrodynamic traps~\cite{Schmidt2005}. 
Here we describe and implement two optical clock state excitation schemes that achieve high-contrast, microsecond-scale operations on large thermal ensembles of ${}^{88}$Sr atoms without a confining potential.

The strongly forbidden $\gnd-\clock$ transition in alkaline-earth-like atoms such as strontium is utilized in a variety of atomic devices, including the most accurate and precise atomic clocks to date~\cite{Bothwell_2019, Mcgrew2018, Brewer2019},
neutral atom quantum computing platforms~\cite{Finkelstein2024,Cao2024,Muniz2025}, and matter-wave interferometers~\cite{Hu2017,rudolf2020,Baynham2026}. 
The bosonic isotope ${}^{88}$Sr typically requires the use of a strong magnetic field to excite the clock transition~\cite{Taichenachev2006}. In fermionic isotopes such as ${}^{87}$Sr, hyperfine mixing enables excitation without an external field. 
In both cases, achievable Rabi frequencies are limited to the kilohertz range with laser intensities on the order of $1\,\text{W}/\text{cm}^{2}$, necessitating atomic confinement or sub-Doppler cooling.

Multiphoton excitation has been discussed as an alternative approach for manipulation of optical clock transitions~\cite{Hong2005, Santra2005, Barker2016, Leanhardt2014}, with recent demonstrations emphasizing applications in atom interferometry, qubit manipulation, and atomic clocks~\cite{Carman2025, He2025, Panelli2025}.
Several of these methods enable higher excitation rates than single-photon excitation at equal laser intensity.
Multiphoton excitation may additionally allow for an orientation of wavevectors that renders the excitation insensitive to atomic velocity, resulting in participation of all atoms in a thermal ensemble~\cite{Grynberg1976, Panelli2025, Glick2025}.
The work presented here combines a Doppler-free wavevector configuration with coherent multiphoton excitation of an optical clock transition.

We demonstrate two Doppler- and recoil-free excitation schemes for the ${}^{88}$Sr $\gnd-\clock$ clock transition. 
With both methods, we show microsecond-scale coherent excitation on thermal ensembles of $3\times10^6$ atoms in free space. 
In the first scheme, three lasers simultaneously drive a coherent three-photon excitation of the transition with $190$\,kHz Rabi frequency and $76\%$ $\pi$-pulse efficiency. 
The excitation lasers are oriented to null the net wavevector, thereby suppressing the Doppler effect.
In the second method, a single-photon excitation from the ground state $\gnd$ to an intermediate state $\ex$ is followed by a two-photon excitation that coherently populates the $\clock$ clock state. 
The sequential excitation achieves an effective $820$\,kHz Rabi frequency with over $90\%$ $\pi$-pulse efficiency while retaining Doppler- and recoil-free features.
With these schemes, we demonstrate a $1000$-fold suppression of Doppler dephasing in a free-space Ramsey sequence, extending the Ramsey coherence time of a $9$\,\textmu K atomic sample from $4.5$\,\textmu s to beyond $4$\,ms. 
Both schemes are widely generalizable to different atomic species and optical clock transitions, and while we have used a bosonic isotope, they could similarly be implemented in fermionic isotopes.

The structure of this paper is as follows. In Sec.~\ref{sec:exp} we provide an overview of the experimental setup. 
The simultaneous and sequential excitation methods are detailed in Secs.~\ref{sec:method1} and \ref{sec:method2}, respectively. 
Experimental characterization of these methods using Ramsey spectroscopy is presented in Sec.~\ref{sec:ramsey}. 
We discuss several applications of these methods in Sec.~\ref{sec:disc} and provide concluding remarks in Sec.~\ref{sec:conc}. We consider systematic effects and provide key derivations in the appendices.

\section{Experiment overview}\label{sec:exp}

\begin{figure}[tbp]
     \centering
    \includegraphics[width=0.9\linewidth]{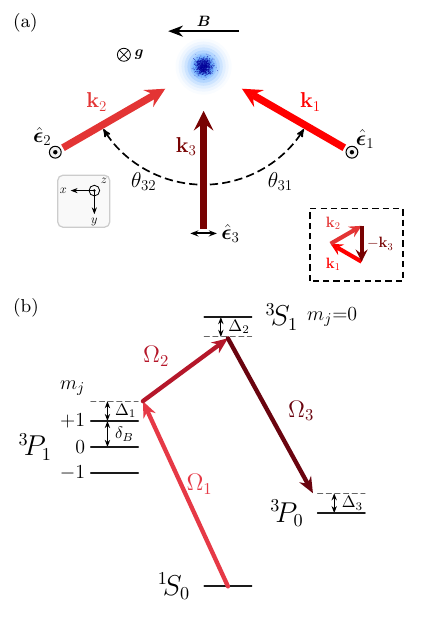}
    \caption{Schematic of Doppler-free (DF) multiphoton excitation of the $\gnd-\clock$ clock transition in alkaline-earth-like atoms.
    (a)~DF configuration of the excitation lasers. The lasers with wavevectors $\mathbf{k}_i$ propagate in a plane orthogonal to gravity $\mathbf{g}$ with lasers 3 and 1 separated by angle $\theta_{31}$ and lasers 3 and 2 separated by $\theta_{32}$. 
    The laser polarization vectors $\hat{\bm{\epsilon}}_i$ (indicated by black dots/arrows) are chosen with respect to an applied magnetic field $\mathbf{B}$ to couple the desired states.
    The DF condition is defined as $\mathbf{k}_{1}+\mathbf{k}_{2}-\mathbf{k}_{3}=\mathbf{0}$ and is illustrated geometrically in the dashed box.
    (b)~Relevant energy levels (not to scale) for multiphoton excitation. Colored arrows indicate lasers coupling atomic states with Rabi frequencies $\Omega_{i}$ and detunings $\Delta_i$, as described in the main text. The Zeeman shift for the $\ex$ manifold is indicated by $\delta_{B}$. }
    \label{fig:3v} 
\end{figure}

As illustrated in Fig.~\ref{fig:3v}(a), three coplanar lasers with wavevectors $\mathbf{k}_{1}, \mathbf{k}_{2}$, and $\mathbf{k}_{3}$ interact with an atomic cloud. 
The wavevectors are oriented in a Doppler-free (DF) configuration where the relative angles $\theta_{31}$ and $\theta_{32}$ are selected so that $\mathbf{k}_{1}+\mathbf{k}_{2}-\mathbf{k}_{3}=\mathbf{0}$. 
This ensures that the first-order Doppler shift $(\mathbf{k}_{1}+\mathbf{k}_{2}-\mathbf{k}_{3})\cdot \mathbf{v}=0$, regardless of the velocity $\mathbf{v}$ of an atom. 
The relevant energy levels for the excitation schemes considered in this work are shown in Fig.~\ref{fig:3v}(b) and are denoted by $\ket{g} \equiv \ket{\gnd}$, $\ket{s_{\pm,0}} \equiv \ket{\ex,\mj{\pm 1,\,0}}$, $\ket{v} \equiv \ket{{}^{3}\!S_{1},\mj{0}}$, and $\ket{e} \equiv \ket{\clock}$, where $\ket{g}$ and $\ket{e}$ are the ground and clock states. 
Not shown in Fig.~\ref{fig:3v}(b) are the ${}^3\!P_2$ and ${}^1\!P_1$ manifolds which we use for repumping and detection and denote by $\ket{d}$ and $\ket{b}$, respectively.

We denote the resonant angular frequencies of the transitions used for excitation as $\omega_{1}, \omega_2$, and $\omega_3$ for the $\ket{g}-\ket{s_+}$, $\ket{s_+}-\ket{v}$, and $\ket{v}-\ket{e}$ transitions, respectively, so that $\omega_1+\omega_2-\omega_3 = \omega_0$, where $\omega_0$ is the $\ket g-\ket e$ transition angular frequency. 
In the lab frame, laser $i$ has angular frequency $\omega_i^\ell$ and is tuned near $\omega_{i}$ for $i\in\{1,2,3\}$.
All excitation lasers are phase-synchronized via beatnote locks to a frequency comb enabling the use of multi-spectral excitation in
phase-sensitive protocols such as Ramsey spectroscopy~\cite{Diddams2020}.

We prepare atomic samples of up to $3\times10^6$ $^{88}$Sr atoms in the ground state $\gnd$ at 7--9\,\textmu K and release the samples into free space for three-photon interrogation.  
A bias magnetic field $\mathbf{B}$ oriented along $\hat{\mathbf{x}}$ induces a Zeeman splitting $\delta_B$ between the $\ex$ sublevels. To achieve the laser couplings shown in Fig.~\ref{fig:3v}(b), lasers 1 and 2 are linearly polarized along $\hat{\mathbf{z}}$ and laser 3 is linearly polarized along $\hat{\mathbf{x}}$. Each coupling has Rabi frequency $\Omega_i$.
All laser $1/\text{e}^2$ intensity radii are much larger than the spatial extent of the atomic cloud (see Appendix~\ref{app:exp}). 
We employ state-selective imaging schemes~\cite{Carman2025} that resolve relative populations in each of the $\ket g\!,\,\ket s\!,\,\ket e\!,$ and $\ket d$ states, where $\ket s$ denotes the entire $\ex$ manifold as the imaging does not distinguish between Zeeman sublevels. Example readout images are shown in the inset of Fig.~\ref{fig:simulrabi} and further experimental details are provided in Appendix~\ref{app:readout}.

\section{Simultaneous three-photon Doppler-free excitation (method 1)}\label{sec:method1}
\subsection{Coherent clock state manipulation and Rabi oscillations}\label{subsec:ex1}

Here all excitation lasers are applied to the atomic sample simultaneously to generate a three-photon coupling between $\ket g$ and $\ket e$.
The angular frequency of laser 1 is given by $\omega_1^\ell=\omega_1 + \Delta_1$, where $\Delta_1$ is the detuning of laser 1 from the $\ket{g}-\ket{s_+}$ resonance. 
For these demonstrations, we set $\Delta_1=2\pi\times 5\,\text{MHz}$ and use a 19\,G bias field to separate the Zeeman sublevels of $\ex$ by $\delta_B=2\pi\times 40\,\text{MHz}$.
The angular frequency of laser 2 is given by $\omega_2^\ell=\omega_2 + \Delta_2-\Delta_1$ where $\Delta_2=2\pi\times400\,\text{MHz}$. 
Lastly, the angular frequency of laser 3 is tuned to $\omega_3^\ell=\omega_3 + \Delta_2 - \Delta_3$, where $\Delta_3= (\omega_1^\ell+\omega_2^\ell-\omega_3^\ell)-\omega_0$ is the net detuning from the clock transition frequency.
The AC Stark shift from the excitation lasers shifts the three-photon resonance by $\delta_\text{AC}$, discussed further in Sec.~\ref{sec:ac1}. 
When tuned to resonance, the three-photon transition occurs with Rabi frequency
\begin{equation}\label{eq:Rabi3v}
    \Omega_\text{3$\nu$} = \frac{\Omega_1 \Omega_2 \Omega_3}{4\Delta_2} \left(\frac{1}{\Delta_1} - \frac{1}{\Delta_1+2\delta_B}\right),
\end{equation}
for our configuration of laser polarizations~\cite{Carman2025}.
\begin{figure}[tbp]
    \centering
    \includegraphics[width=\linewidth]{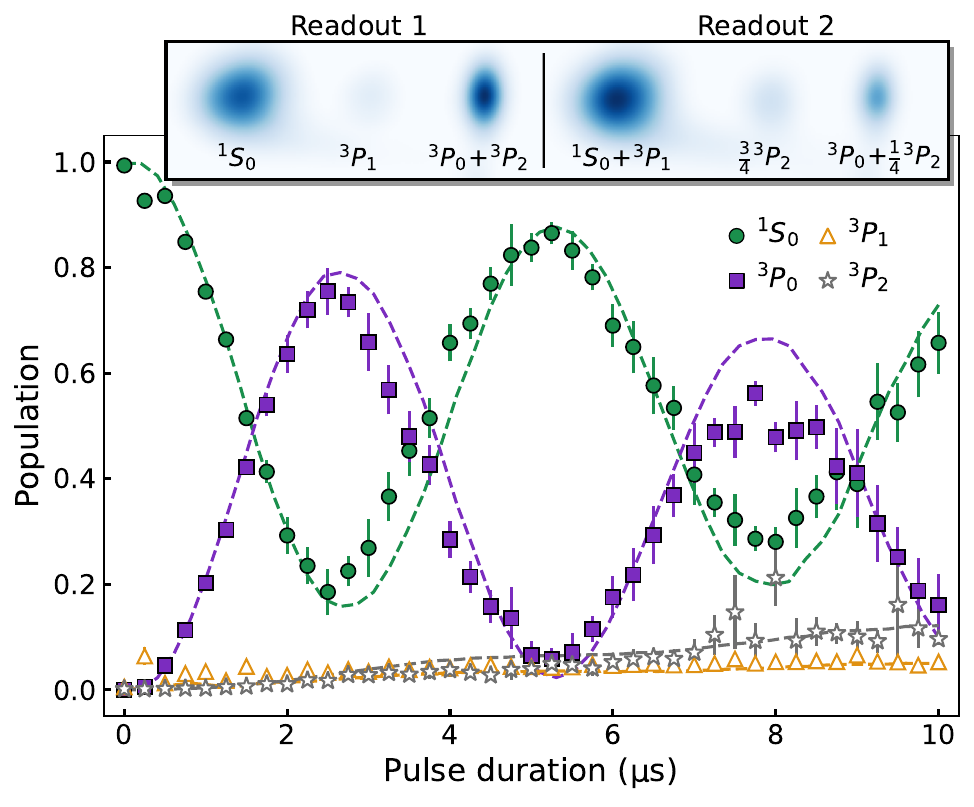}
    \caption{Three-photon Rabi oscillations on the ${}^{88}$Sr $\gnd-\clock$ clock transition. Each data point is the mean of five measurements with error bars given by the standard deviation.
    Circles (green), squares (purple), triangles (orange), and stars (gray) indicate populations in $\gnd$, $\clock$, $\ex$, and ${}^{3}\!P_{2}$, respectively.  
    Dashed curves show corresponding simulated population dynamics. 
    The measured Rabi frequency is $192(3)\,$kHz. 
    The inset shows an example of fluorescence images for the two clock state readout schemes combined to estimate relative populations.}
    \label{fig:simulrabi}
\end{figure}

We set laser intensities corresponding to measured couplings $\Omega_1=2\pi\times3.82(2)$\,MHz and $\Omega_2\Omega_3/2\Delta_2= 2\pi\times0.59(1)\,$MHz.
We directly measure $\Omega_1$ from Rabi oscillations between $\ket g$ and $\ket{s_+}$, while we measure the product of Rabi frequencies for lasers 2 and 3 via two-photon (Raman) Rabi oscillations between $\ket{s_+}$ and $\ket e$. 
The number in parentheses is the standard error of the Rabi frequency estimate.
We estimate the individual Rabi frequencies for lasers 2 and 3 to be $\Omega_2=2\pi\times23$\,MHz and $\Omega_3=2\pi\times22$\,MHz from laser power and beam size measurements. 
From these values we infer a two-photon Rabi frequency within 10\% of the measured value.
At these excitation parameters, we keep the single-photon detunings much larger than the Doppler width of the atomic ensemble ($\Delta_D\approx 2\pi \times100$\,kHz) to mitigate the effect of non-uniform three-photon Rabi frequency and AC Stark shift across different velocity classes~\cite{Grynberg1976}.
We discuss excitation limitations due to velocity-dependent coupling strength in Appendix~\ref{app:sys_finite_detuning}.

On-resonance three-photon excitation leads to Rabi oscillations in the population of $\ket g$ and $\ket e$ as shown in Fig.~\ref{fig:simulrabi}, in good agreement with simulation (see Appendix~\ref{app:sim_model}). 
We measure a three-photon Rabi frequency of $192(3)$\,kHz, which deviates from the value predicted from Eq.~\eqref{eq:Rabi3v} by $10\%$ using the measured single-photon and Raman Rabi frequencies. This reduction from the theoretical value is primarily attributed to intensity inhomogeneities across the cloud.
We measure $76(2)\%$ clock state excitation fraction at the Rabi oscillation peak with $\approx\!5\%$ population of intermediate states at the $\pi$-pulse time.

We minimize off-resonant scattering from the individual intermediate transitions by ensuring $\Delta_2\gg\Omega_2,\Omega_3$, which allows for observation of high-contrast Rabi oscillations. 
Off-resonant scattering details are discussed and calculated in Appendix~\ref{subsec:scat}.
We operate with $\Delta_1\approx\Omega_1$ to increase the three-photon Rabi frequency with fixed laser power. 
Error-function-shaped pulses suppress oscillations in the intermediate-state populations.
We provide details of the simulation model in Appendix~\ref{app:sim_model} and a budget for all contributing mechanisms on the achievable $\pi$-pulse excitation fraction in Table~\ref{tab:fid}.

\subsection{Doppler-free resonance}\label{sec:dfres}
We confirm the DF behavior of the three-photon excitation method by comparing the spectroscopic linewidth of the three-photon $\ket g -\ket e$ transition to Doppler-sensitive single-photon excitation of the $\ket{g}-\ket{s_+}$ transition using Rabi spectroscopy~\cite{Ramsey1986}. 
These transitions have approximately equal single-photon Doppler sensitivity in strontium, allowing for direct comparison.

\begin{figure}[tbp]
    \centering
    \includegraphics[width=\linewidth]{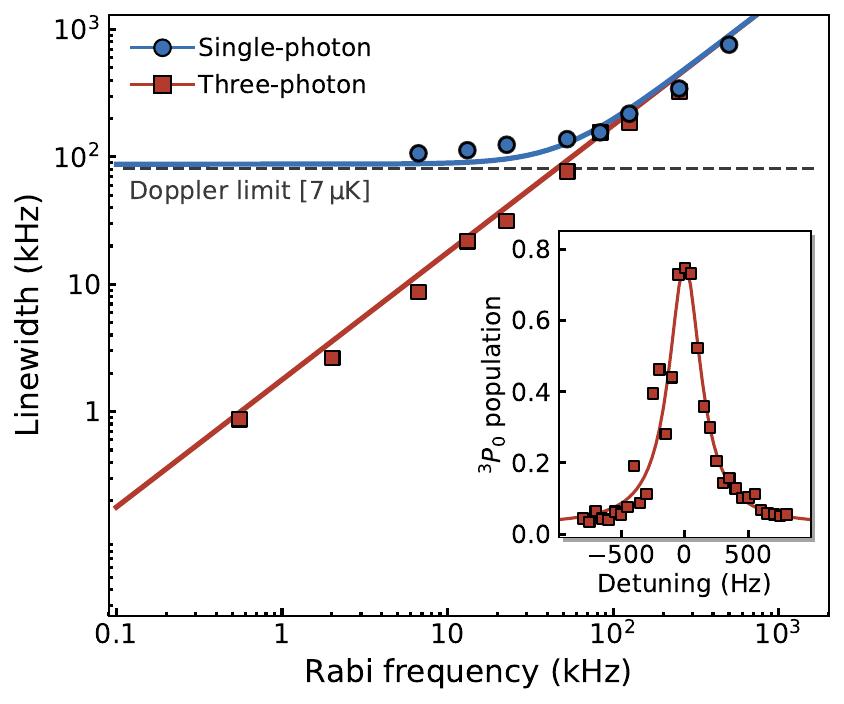}
    \caption{Three-photon suppression of Doppler broadening. Data points are FWHM linewidths of the single-photon (blue) and three-photon (red) resonances as a function of Rabi frequency. Solid lines are theory for Fourier (red) and Doppler (blue) linewidth limits for a $7$\,\textmu K sample. 
    Single-photon excitation linewidth is bounded by the Doppler limit of the optical transition while three-photon excitation achieves Fourier-limited linewidths.
    The inset shows a DF resonance feature with a $290(20)$\,Hz linewidth with $75\%$ $\clock$ excitation fraction.}
    \label{fig:simulline}
\end{figure}

We measure the resonances by interrogating the atomic sample with the excitation lasers for a pulse duration set to the $\pi$-pulse time at a given laser power. We measure atomic state population as a function of the frequency of laser 1 and fit to find the full-width-at-half-maximum (FWHM) linewidth. We vary the excitation Rabi frequencies by adjusting laser power.
By decreasing the three-photon Rabi frequency, we are able to achieve resonance features far below the single-photon Doppler width of the atomic ensemble. 
The linewidth of the three-photon resonance scales with the inverse of the excitation pulse duration as shown in Fig.~\ref{fig:simulline} due to the Fourier-limited linewidth of a square $\pi$-pulse $\delta \nu_\text{F} \approx 0.886/T_\pi$.
This is not the case for single-photon excitation, which saturates at the Doppler limit.
We measure three-photon resonance features with FWHM below $300$\,Hz while maintaining $75\%$ clock state excitation fraction, shown in the inset of Fig.~\ref{fig:simulline}. 
Further reduction of the spectroscopic linewidth is limited both by residual Doppler broadening from imperfect alignment of the excitation lasers and by atomic free-fall out of the excitation and readout region.

In this low Rabi frequency regime, we measure a high excitation fraction and exceptionally narrow linewidths on an optical clock transition using free-space thermal ensembles.  
Application of three-photon clock state excitation to timekeeping has previously been proposed~\cite{Hong2005, Carman2025}.
For a resonance of FWHM linewidth $\delta\nu$ at the clock frequency $\nu_0$, interrogated with $N$ atoms at the quantum projection noise (QPN) limit, the single-shot statistical frequency instability is of order $\delta\nu/(\pi\nu_0\sqrt{N})$~\cite{Santarelli1999}. 
The narrow-line feature in Fig.~\ref{fig:simulline} and $N=3\times10^6$ atoms leads to a potential QPN-limited fractional frequency instability of $\sigma_y(\tau)=10^{-16}/\sqrt{\tau}$ for a 1\,s cycle time and averaging time $\tau$ in seconds.
This can be improved with further reduction of the residual wavevector along with longer interrogation times and higher repetition rates.

\subsection{Three-photon AC Stark shifts}\label{sec:ac1}
Operating in free space eliminates shifts from confining potentials and the associated need to utilize magic wavelength traps~\cite{Katori2003}, though multi-spectral excitation introduces additional sources of AC Stark shift to be considered. 
The net AC Stark shift on the $\ket g - \ket e$ transition from the excitation lasers is approximately
\begin{equation}
    \delta_\text{AC}=-\frac{\Omega_1^2}{2}\frac{\Delta_1+\delta_B}{\Delta_1(\Delta_1+2\delta_B)} + \frac{\Omega_3^2}{4\Delta_2} - \beta I_2 ,
\end{equation}
where $I_2$ is the peak intensity of laser 2 and $\beta=70\,\text{Hz}\,/(\text{W}/\text{cm}^{2})$ is computed with the differential scalar polarizability between the $\ket g$ and $\ket e$ levels at the wavelength of laser 2~\cite{2025Portal}.
In this work, lasers 1, 2, and 3 contribute $-40$\,kHz/mW, 5\,Hz/mW, and 80\,kHz/mW, respectively, to the AC Stark shift.

The narrow-line resonance shown in the inset of Fig.~\ref{fig:simulline} was measured using $\approx$10\,\textmu W of power in lasers 1 and 3 and 11\,mW in laser 2.
In that demonstration, the net AC Stark shift contains a contribution of $-400\,$Hz from laser 1, $60$\,Hz from laser 2, and $800\,$Hz from laser 3. These scalings suggest that in future work the AC Stark shift can be reduced further with optimized laser intensities.

\section{Sequential Doppler-free excitation (method 2)}\label{sec:method2}

\subsection{Sub-microsecond excitation of the clock transition}\label{subsec:ex2}

In this method, we manipulate the clock transition via sequential, time-separated resonant drives of the $\ket g - \ket{s_+}$ single-photon transition using laser 1 and the $\ket{s_+} - \ket e$ Raman transition with intermediate state $\ket v$ using lasers 2 and 3.
For high-contrast excitation, all excitation pulse durations and the duration between pulses must be much shorter than the lifetime of the intermediate state $\ket {s_+}$ to minimize spontaneous decay.

The lasers remain in the DF configuration from method 1, so that $\mathbf{k}_1$ and the effective wavevector for the Raman process $\mathbf{k}_R \equiv \mathbf{k}_2-\mathbf{k}_3$ are equal and opposite, i.e., $\mathbf{k}_1 + \mathbf{k}_R = \mathbf{0}$.
The frequency and power of laser 1 are adjusted to be on resonance with the $\ket g - \ket{s_{+}}$ transition with $\Omega_1 = 2\pi\times 4.63(3)\,$MHz.
We relabel the single-photon detuning $\delta_1 \equiv \omega_1^\ell - \omega_1$ to distinguish the notation between the two methods.
The frequency of laser 2 is adjusted so that the $\ket{s_+}-\ket e$ Raman transition is resonant using an intermediate-state detuning of $\Delta_2=2\pi \times 400\,\text{MHz}$. Here $\Delta_2$ is the detuning of laser 2 from the $\ket{s_+} -\ket v$ resonance, independent of the frequency of laser 1.
Thus, $\omega_2^\ell = \omega_2+\Delta_2$ and $\omega_3^\ell = \omega_3+\Delta_2-\delta_R$ where we have defined a Raman detuning $\delta_R = (\omega_2^\ell-\omega_3^\ell) - (\omega_2-\omega_3)$. Both $\delta_1$ and $\delta_R$ are nominally zero for this excitation method.
The power of laser 3 is adjusted so that the Raman transition has a Rabi frequency of $\Omega_R = \Omega_2\Omega_3/2\Delta_2 = 2\pi\times1.12(2)$\,MHz.
The Rabi frequencies $\Omega_1$ and $\Omega_R$ are obtained directly from the measured on-resonance Rabi oscillation frequency for their respective transitions.
All remaining laser parameters are unchanged from method 1 (see Sec.~\ref{sec:method1}).

\begin{figure}[tbp]
    \centering
    \includegraphics[width=\linewidth]{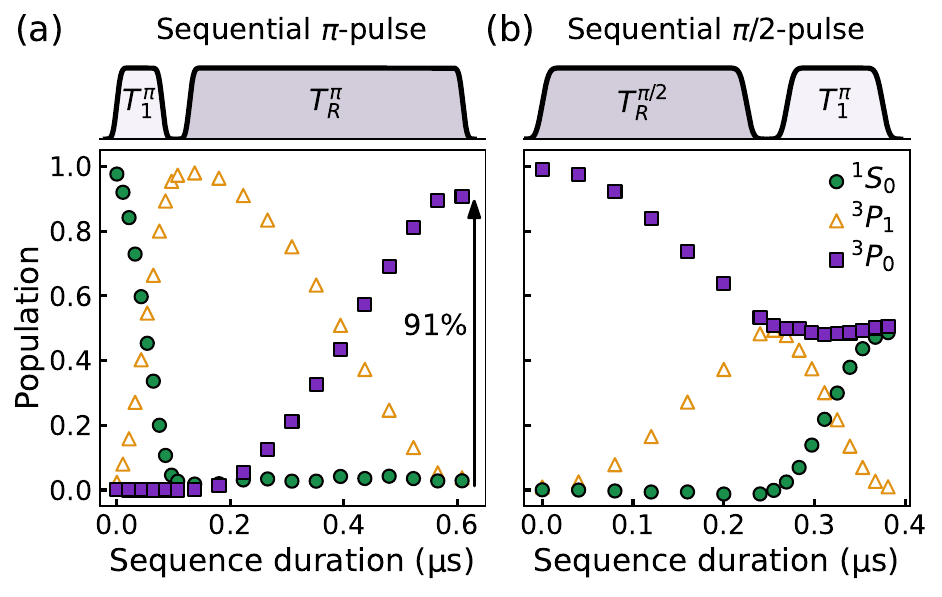}  
    \caption{Sequential DF excitation of the $\gnd-\clock$ clock transition.
    (a) Sequential $\pi$-pulse for coherent $\clock$ excitation. Measured populations for $\clock$, $\ex$, and $\gnd$ are indicated by squares (purple), triangles (orange), and circles (green), respectively.
    The effective Rabi frequency is $820$\,kHz and a maximum excitation fraction of $91(1)\%$ is observed. 
    Note that the $\gnd$ readout port contains leakage population from $\ex$ decay during imaging (see Appendix~\ref{app:exp}).
    (b) Sequential $\pi/2$-pulse preparing an equal superposition of the ground and clock states in under $400\,$ns. Population of ${}^3\!P_2$ is measured to be below $3\%$ and is omitted for visual clarity.}
    \label{fig:seq}
\end{figure}

We show coherent clock state excitation with greater than $90\%$ population transfer in $610$\,ns in Fig.~\ref{fig:seq}(a).
The atomic ensemble is prepared in $\ket g$, then laser 1 is applied for a time $T_1^\pi = \pi/\Omega_1$ to transfer atoms to $\ket{s_+}$. 
Immediately after, lasers 2 and 3 are applied for a time $T_R^\pi=\pi/\Omega_R$ and population is transferred to $\ket e$. 
The effective Rabi frequency is determined by the total duration of the pulse sequence, 
corresponding to an 820\,kHz Rabi frequency.
The residual excitation of the $\ket d$ manifold does not exceed 3\%.
Additionally, the excitation imparts zero net photon recoil.

We also demonstrate an effective $\pi/2$-pulse, creating a $\ket g + \ket e$ superposition state (unnormalized) in $380\,$ns as shown in Fig.~\ref{fig:seq}(b). 
We begin with the ensemble in the clock state $\ket e$ (see Appendix~\ref{app:prep}) to minimize the duration of the $\pi/2$-pulse sequence.
We apply the Raman drive for a duration $T_R^{\pi/2} =\pi/(2\Omega_R)$ to create a superposition state $\ket{s_+}+\ket{e}$ then
apply the laser 1 drive for a duration $T_1^\pi$ to coherently transfer $\ket{s_+}$ population to $\ket{g}$. 
This results in a final ground-clock state superposition $\ket g +\ket e$.

Any spontaneous emission while in the $\ket{s_+}$ state or incomplete Raman transfer results in loss of coherence and limits the maximum excitation fraction.
In strontium, the $\ket{s_+}$ state has a $21$\,\textmu s natural lifetime. 
Population decay during the pulse sequence limits the maximum excitation fraction to $98.6\%$. This limit is not fundamental and can be improved with increased Rabi frequencies.
The measured excitation fraction is additionally lowered by spontaneous Raman scattering and laser intensity inhomogeneity across the cloud,
which result in $2$--$3\%$ reduction in peak excitation fraction each. All other mechanisms contribute less than $0.1\%$ to the decrease in peak excitation. These contributions are summarized in Table~\ref{tab:fid}.

For comparison, direct single-photon clock state excitation in ${}^{88}$Sr has a coupling that scales $\propto B\sqrt{I}$~\cite{Taichenachev2006}. With the magnetic fields and intensities used in this work, one would achieve a Rabi frequency of order $30$\,Hz. Improving this to the coupling rates we have demonstrated above would require a $2.5\times 10^4$ relative increase in $B\sqrt{I}$, incompatible with realistic magnetic fields and laser intensities. 
This method provides a faster alternative to STIRAP-like approaches for coherent population transfer~\cite{He2025, Bergmann1998}.
Similar excitation schemes have been shown in solid-state systems such as superconducting and nuclear spin qudits using microwave transitions~\cite{Godfrin2018, Neeley2009}.  
Here we have shown applicability to optical clock transitions in neutral atom systems.
In Sec.~\ref{sec:ramsey} below we demonstrate suitability for phase-sensitive measurements such as Ramsey spectroscopy. 

\subsection{Doppler-free phase dynamics for sequential excitation}\label{subsec:ex2phase}

Although the sequential excitation method consists of individually velocity-sensitive pulses, it retains robustness to Doppler dephasing similar to that of method 1 upon completion of the excitation.
We analytically solve a simplified three-level system (see Appendix~\ref{app:3vseq}) for the relative phase of the superposition state $\ket{g}+e^{i\varphi(\tau)}\ket{e}$ created by a sequential $\pi/2$-pulse starting in the clock state and show that the wavevector geometry enables DF evolution with a small but static phase spread due to finite pulse durations. 

The phase at time $\tau$ after the sequential excitation is given by
\begin{equation}\label{eq:seq_open}
    \varphi(\tau) = (\delta_1 + \delta_R)\tau+ \delta_1\frac{T_1^\pi}{2} + \delta_R\left(T_1^\pi  + T_s + \frac{1}{\Omega_R}\right),
\end{equation}
where $T_s$ is the time between the sequential excitation pulses. 
The individual detunings are Doppler shifted by $\delta_i \to  \delta_i -\mathbf{k}_i\cdot\mathbf{v}$.
Thus, under the DF condition $\mathbf{k}_R=-\mathbf{k}_1$, 
the first term in Eq.~\eqref{eq:seq_open} is velocity-independent.
The only velocity-dependent phase that remains is a static offset determined by the finite pulse durations and the inter-pulse gap, resulting in an ensemble phase spread of
\begin{equation}\label{eq:open_spread}
    \sigma_\varphi = |\mathbf{k}_1|\sigma_v\left(  \frac{T_1^\pi}{2} + T_s + \frac{1}{\Omega_R}   \right),
\end{equation}
where $\sigma_v$ is the one-dimensional velocity spread of the ensemble in the direction of $\mathbf{k}_1$.
Crucially, this has no dependence on $\tau$ and the $\ket g-\ket e$ coherence from the sequential excitation is Doppler-free.
For the excitation parameters outlined above with an inter-pulse gap of $T_s = 50$\,ns and a $9$\,\textmu K atomic sample, we estimate a phase spread of $\sigma_\varphi\approx 70$\,mrad.

\subsection{Raman AC Stark shifts}

The dominant AC Stark shifts in the sequential method are from the Raman lasers shifting the $\ket{s_+} - \ket e$ two-photon resonance by
\begin{equation}\label{eq:AC2}
    \delta_\text{AC,2}=\frac{\Omega_2^2-\Omega_3^2}{4\Delta_2}.
\end{equation}
For our excitation parameters, the resonance shifts by $400$\,kHz from laser 2 and $-950$\,kHz from laser 3. We tune to this shifted resonance when performing the Raman excitation.
The net shift can be nulled with an appropriate balance of Raman laser intensities.
The remaining shifts are much smaller than the relevant Rabi frequencies and can be neglected. The $\sigma^{-}$ component of laser 1 off-resonantly couples $\ket{g}-\ket{s_-}$ and shifts the ground state by $70$\,kHz, while additional far-off-resonant contributions all remain at the $100$\,Hz-level or below.

Because the atoms are interrogated in free space, AC Stark shifts are only present while the excitation lasers are on (under a microsecond), and therefore lead to negligible effects on excitation fraction and only a small, static phase imprint set by the net shift and pulse durations.
The inhomogeneity of $ \delta_\text{AC,2}$ across the atomic sample has a $20$\,kHz standard deviation, much smaller than $\Omega_R$. It produces a static phase spread on the order of $10$\,mrad, comparable to the finite-pulse-duration spread of Eq.~\eqref{eq:open_spread}, and similarly does not accrue during the dark time.

\section{Ramsey spectroscopy}\label{sec:ramsey}
In the previous sections, both methods 1 and 2 show clock state population transfer with high excitation fraction. 
Ramsey spectroscopy additionally requires that the $\ket g-\ket e$ phase coherence is maintained after initial excitation. 
The wavevector geometry is designed to suppress Doppler dephasing.
Remaining velocity sensitivity originates from the uncanceled net wavevector $\Delta\mathbf{k} = \mathbf{k}_1 + \mathbf{k}_2 - \mathbf{k}_3$ due to imperfect alignment. 
Here we describe the impact of this residual Doppler sensitivity and finite pulse times on phase inhomogeneity in Ramsey spectroscopy. 
We demonstrate $1000$-fold coherence-time enhancement compared to a Doppler-sensitive excitation.

\subsection{Doppler-free Ramsey sequences}

The Ramsey sequences for both excitation methods consist of opening and closing effective $\pi/2$-pulses separated by a dark (sensor) time $\tau$. The opening sequence creates a superposition between the ground and clock states $\ket{g}+e^{i\phi_A}\ket{e}$ with a relative phase $\phi_A$.
During the dark time, the state accrues an additional relative phase shift $\phi_s$ and the phase of laser 1 is shifted by a programmed $\phi_1$. 
The closing pulse sequence imprints a phase $\phi_B$ and converts the total accumulated phase to a ground state population given by
\begin{equation}
    P_g = \frac{1}{2}\Big (1-C\cos(\Phi)\Big ) ,
\end{equation}
where $C$ is the contrast and $\Phi = \phi_A+\phi_s+\phi_1 +\phi_B$ is the overall Ramsey phase. 
Varying $\phi_1$ at a fixed $\tau$ traces out Ramsey fringes in atomic state populations. 
The phase from the opening and closing sequences $\phi_A+\phi_B$ is method-dependent and discussed below.

For both excitation methods, the sensing phase for an atom with velocity $\mathbf{v}$ is given by
\begin{equation}\label{eq:phi_sense}
    \phi_s \approx (\omega_1^\ell+\omega_2^\ell-\omega_3^\ell-\omega_0)\tau + \Delta\mathbf{k}\!\cdot\!\mathbf{v}\tau .
\end{equation}
For an ensemble of atoms, the residual uncanceled wavevector leads to Doppler dephasing at a rate of $|\Delta\mathbf{k}|\sigma_v$.
This also leads to residual recoil and a spatial displacement of $\hbar|\Delta\mathbf{k}|\tau/m$ between the atomic states, where $m$ is the mass of a ${}^{88}$Sr atom.

Method 1 is a direct three-photon drive on the $\ket g - \ket e$ transition and so can be modeled as an effective two-level system.
The residual Doppler shift $\Delta\mathbf{k}\cdot\mathbf{v}$ leads to a Bloch vector tilt of $\tan^{-1} (\Delta\mathbf{k}\cdot\mathbf{v}/\Omega_{3\nu})\approx\Delta\mathbf{k}\cdot\mathbf{v}/\Omega_{3\nu}$~\cite{steck}.
Thus, the phase spreads of the opening and closing Ramsey pulses are velocity-dependent when $|\Delta\mathbf{k}|>0$ with resulting widths $\sigma_{\phi_A} = \sigma_{\phi_B} = |\Delta\mathbf{k}|\sigma_v/\Omega_{3\nu}$.
For milliradian-level laser alignment accuracy and 200\,kHz three-photon Rabi frequency, this phase spread is below 1\,mrad in our system. 
Thus, the phase contributions from the opening and closing pulses are static and nearly homogeneous.

The Ramsey sequence for excitation method 2 consists of two sequential $\pi/2$-pulses. 
The opening $\pi/2$-pulse consists of a Raman $\pi/2$-pulse followed by a single-photon $\pi$-pulse, leaving a coherent superposition $\ket g + e^{i \phi_A}\ket e$. 
The closing pulse sequence consists of a Raman $\pi$-pulse followed by a single-photon $\pi/2$-pulse. 
This maps the Ramsey phase $\Phi$ onto the relative population between $\ket g$ and $\ket{s_+}$ before readout. 
The opening and closing pulses contribute a total phase given by (see Appendix~\ref{app:3vseq})

\begin{equation}
\begin{aligned}\label{eq:phifp_seq}
    \phi_A + \phi_B =&\, \delta_1\frac{T_1^\pi}{2} + \delta_R\left(T_1^\pi  + T_s + \frac{1}{\Omega_R}\right) \\
    & + \delta_R\frac{T_R^\pi}{2} + \delta_1\left(T_R^\pi  + T_s + \frac{1}{\Omega_1}\right) .
\end{aligned}
\end{equation}
This is velocity-dependent via $\delta_i\to\delta_i-\mathbf{k}_i\cdot\mathbf{v}$.
Since $\mathbf{k}_1\approx-\mathbf{k}_R$, the closing pulse partially cancels the phase from the opening pulse. 
Full cancellation can be achieved by choosing $\Omega_1=\Omega_R$.
In this work, we estimate a residual static phase spread $\sigma_{\phi_A+\phi_B}$ under 25\,mrad.

\begin{figure*}[tbp]
    \centering
    \includegraphics[width=\linewidth]{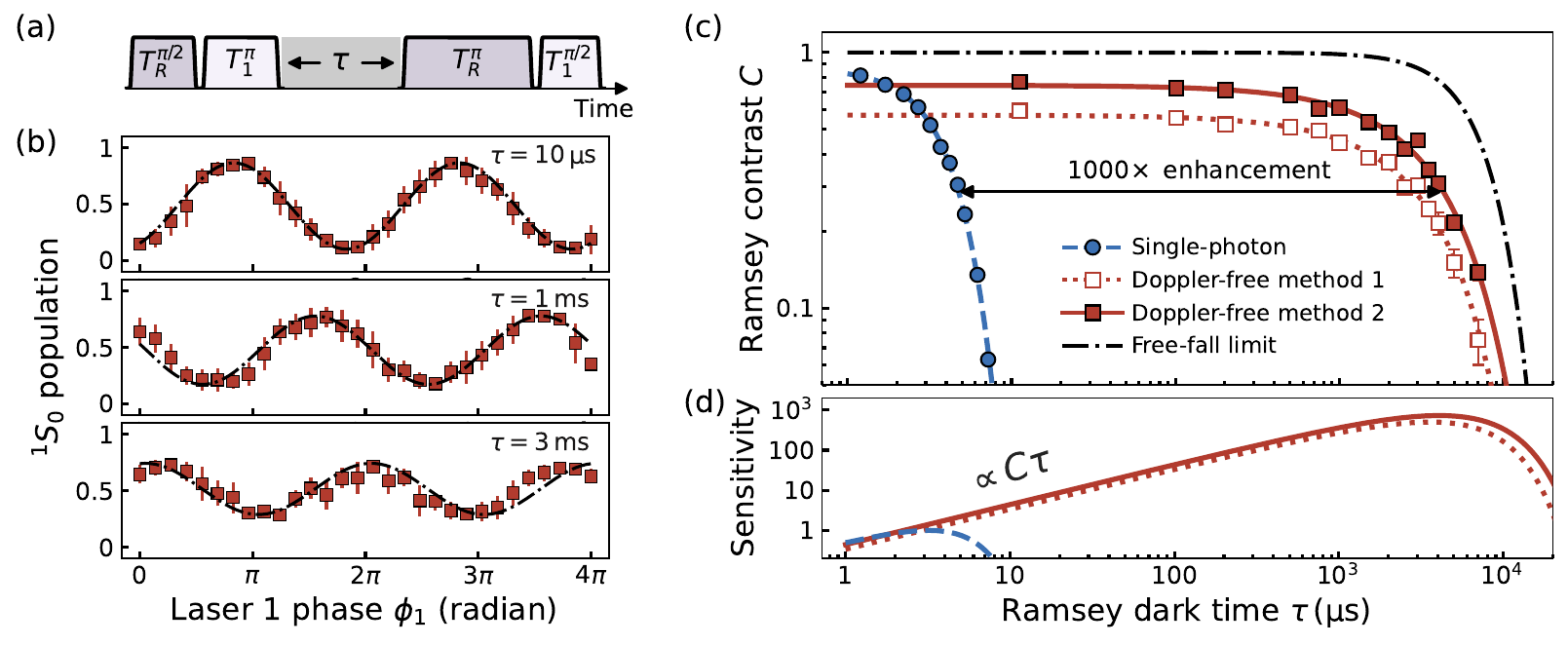}
    \caption{Ramsey fringes and coherence time comparisons between single-photon and Doppler-free methods 1 and 2. (a) Pulse sequence for the sequential DF method 2. The pulse sequence for single-photon and DF method 1 is the conventional form of two time-separated $\pi/2$-pulses. (b) Ramsey fringes for dark times $\tau$ of $10$\,\textmu s (top), $1$\,ms (middle), and $3$\,ms (bottom) using sequential DF method. 
    The closing phase of laser 1 is varied to trace out a phase-dependent ground state population at the end of the Ramsey sequence. 
    Each data point is the mean of ten repeated measurements and error bars are standard deviations.
    Dashed black curves are sinusoidal fits to the data used to estimate the contrast.
    We attribute the shortened Ramsey fringe period at large dark times to slow drift of our optical frequency reference during the experiment run time (see Appendix~\ref{app:lasers}).
    (c) Comparison of Ramsey sequence contrast as a function of $\tau$ for
    single-photon (blue circles), simultaneous DF (open red squares), and sequential DF (filled red squares) excitation. 
    The data points are from sinusoidal fits to Ramsey fringe measurements and error bars indicate the standard error of the contrast estimate. 
    Lines are corresponding fits to the contrast decay data using a product of Gaussian and exponential decay to capture inhomogeneous and homogeneous decay effects, respectively (see Appendix~\ref{app:fits}).
    A factor of $1000$ coherence-time enhancement is shown between multiphoton and single-photon excitation.
    The dashed black line is the limit of measurable contrast enhancement for the sequential method due to ensemble free-fall out of the excitation beam waist and thermal expansion.
    (d) Relative sensitivity ($\propto C\tau$) for single-photon and DF excitation methods. Lines indicate the sensitivity scaling for each method normalized to the peak single-photon value.}
    \label{fig:ramseydata}
\end{figure*}

The recoil from the time separation between the laser pulses leads to a spatial separation between the states of $\hbar k_R T_s/m= 3\,$\AA{} and is much less than the coherence length.
The displacement $\hbar|\Delta\mathbf{k}|\tau/m$ between states due to residual recoil from uncanceled net wavevector dominates the total displacement at the millisecond-scale dark times in the demonstrations below.

\subsection{Demonstration of coherence-time enhancement}

Here we measure the Ramsey coherence times for each excitation scheme. We use the parameters from Sec.~\ref{subsec:ex1} and Sec.~\ref{subsec:ex2} for methods 1 and 2, respectively. 
We compare Ramsey contrast decay using these methods to a single-photon Ramsey experiment with the $\ket{g}-\ket{s_+}$ transition.

We measure Ramsey contrast $C$ at fixed $\tau$ by scanning the closing $\pi/2$-pulse phase $\phi_1$ over $4\pi$ radians and measuring the relative state populations. We then fit the function $P_g(\phi_1)=A-\frac{C}{2}\cos(\phi_1+\phi_{0})$ to the data, with $A$ and $\phi_0$ as free parameters.
Note that this is a conservative estimate for the contrast due to phase noise in our system.
Example phase scans for dark times $\tau=\{10\,\text{\textmu s},\, 1\,\text{ms},\, 3\,\text{ms}\}$ are shown in Fig.~\ref{fig:ramseydata}(b). 
We plot the Ramsey contrast as a function of $\tau$ for both DF schemes and the single-photon control experiment in Fig.~\ref{fig:ramseydata}(c).
For single-photon excitation, 
the Ramsey contrast rapidly decays due to Doppler dephasing, resulting in a measured $1/\text{e}$ contrast decay time of $4.5(2)$\,\textmu s.
The DF techniques suppress this dephasing $1000$-fold, extending the contrast decay time to $4.5(4)$\,ms for the sequential method and $4.1(4)$\,ms for the simultaneous method.
We maintain similar levels of contrast at short interrogation times in the DF and single-photon Ramsey sequences.
When applied to a theoretical sensor with a sensitivity that scales with the product of contrast and dark time, this results in a factor of $730$ improvement in sensitivity, as shown in Fig.~\ref{fig:ramseydata}(d).  

The dominant contribution to contrast loss stems from residual Doppler dephasing from the uncanceled net wavevector. We estimate a residual wavevector magnitude of $|\Delta\mathbf{k}|/|\mathbf{k}_1| = 10^{-3}$, consistent with $\leq\!1$\,mrad of cumulative laser misalignment (see Appendix~\ref{app:sys_angle}). 
An additional mechanism contributing to contrast loss is attributed to the free fall and thermal expansion of the atomic ensemble out of the waist of the excitation lasers for large $\tau$. 
We estimate this effect to limit the achievable $1/\text{e}$ contrast decay time to $9.4$\,ms ($9.7$\,ms) for the simultaneous (sequential) DF methods (see Appendix~\ref{app:sys_freefall}).
However, this is not a fundamental limit and can be improved with larger beam sizes or by operating in an atomic fountain configuration~\cite{Santarelli1999}.

\section{Discussion}\label{sec:disc}
The methods presented here are applicable to a wide variety of platforms for quantum sensing, simulation, and information processing where performance could benefit from increased atom number during operation, insensitivity to sample temperature, elimination of photon recoil, or fast excitation.
All of these are enabled by multiphoton clock state excitation with a DF wavevector geometry.
Here we discuss several applications.

\paragraph{Quantum-enhanced sensing.}
Both methods show promise for devices aiming to achieve sensitivities beyond the standard quantum limit.
The achievable enhancement in sensitivity relative to classical coherent spin states scales as $\sqrt{N}$, where $N$ is the number of atoms~\cite{Kitagawa1993, Wineland1994, Pezze2018, Giovannetti2004}. 
DF excitation enables high-contrast clock state manipulation in thermal ensembles, leading to a substantial atom-number advantage compared to single-photon excitation schemes and the potential for significant metrological gain with reduced cooling and confinement requirements.
Furthermore, quantum enhancement in atomic devices is often achieved using atom-light interactions in optical cavities~\cite{Hosten2016, Greve2022, Yang2025}.  
DF excitation techniques enable excitation orthogonal to the cavity axis.
This provides flexibility in mirror coating and geometry requirements for squeezing cavities.

\paragraph{Precision metrology.}
Rapid generation and manipulation of spin-squeezed states in large, free-space atomic ensembles using DF excitation has direct relevance to atom-interferometric gravitational wave detection~\cite{Dimopoulos2008, Abe_2021}. 
These systems plan to use large-momentum-transfer techniques with the optical clock transition in strontium and rely on spin-squeezing to reach the necessary sensitivity for terrestrial experiments.
   
\paragraph{Optical tweezer arrays and Rydberg excitation.}

The reduction of the net photon recoil from the simultaneous three-photon excitation method allows for shallower potentials to be used for Lamb-Dicke confinement, lowering power requirements and AC Stark shift effects for systems like optical tweezer arrays~\cite{Norcia2019, Hong2005}.
Additionally, accessing single-photon Rydberg transitions from the $\clock$ clock state (and the similar metastable ${}^{3}\!P_{2}$ state) in alkaline-earth-like atoms is common practice~\cite{EndresRydberg, KaufmanRydberg}. 
The excitation schemes presented here may facilitate and accelerate the preparation of these states.

\section{Conclusion}\label{sec:conc}
In conclusion, we have demonstrated two excitation schemes that achieve microsecond-scale coherent control of the strongly forbidden $\gnd-\clock$ clock transition in $^{88}$Sr. 
These methods allow for high-contrast excitation in large thermal ensembles without a confining potential and suppress Doppler dephasing effects by three orders of magnitude.
The techniques shown here are applicable to narrow-line transitions in many atomic species and show promise for a wide variety of quantum devices.

\begin{acknowledgments}
We wish to thank Jan Rudolf and Joonhee Choi for helpful conversations.
This work was supported by the Gordon and Betty Moore Foundation Grant No. 7945, the National Science Foundation QLCI Award No. 1559523, and the U.S. Department of Energy Office of Science National Quantum Information Science Research Centers, as part of the Q-NEXT center.
\end{acknowledgments}

\section*{Data availability}
The data that support the findings of this article are openly available~\cite{datalink}.

\appendix

\section{Methods}\label{app:exp}

\subsection{Sample preparation}\label{app:prep}
We prepare the atomic samples of $^{88}$Sr atoms in the ground state $\gnd$ at $7$--$9$\,\textmu K with a dual-stage magneto-optical trap (MOT)~\cite{Katori1999, ludlow2008strontium}. 
The MOT is overlapped with an $813$\,nm optical lattice oriented along the direction of gravity ($\hat z$ in Fig.~\ref{fig:3v}) with a depth of $40$\,\textmu K. 
Upon termination of the MOT, the atoms
remain confined in the lattice while a $19$\,G bias field is ramped on in $15$\,ms. 
This field separates the Zeeman sublevels of the intermediate states $\ex$ and ${}^3\!S_1$ and provides a well-defined quantization axis. 
The lattice is then terminated, and the atomic sample is released into free space for three-photon interrogation.
Upon free-space release, the atoms occupy a spatial extent of $40$\,\textmu m ($130$\,\textmu m) standard deviation in the transverse (axial) direction and undergo free-space expansion.

To initialize the ensemble in the clock state $\clock$, prior to turning off the lattice light we perform a sequential excitation using the procedure described in Sec.~\ref{sec:method2}. This transfers $\approx 90\%$ of the $\gnd$ atoms to the $\clock$ state.
We then use a $2$\,ms pulse of light resonant with the $461$\,nm $\gnd-{}^1\!P_1$ transition to clear out any atoms that were unsuccessfully transferred.

\subsection{Excitation lasers}\label{app:lasers}

All excitation lasers are injection-locked diode lasers seeded by external cavity diode lasers (ECDLs) to amplify the optical power delivered to the atomic sample.
The $689$\,nm ECDL is stabilized using an ultra-low-expansion (ULE) reference cavity with a finesse of $3\times10^5$ and 5\,kHz linewidth (Stable Laser Systems) using the Pound-Drever-Hall method~\cite{Drever1983}. 
This stabilized light has a fractional frequency deviation $<\!10^{-15}$ at 1\,s and a stability of $<\!1$\,Hz.
The repetition rate of a frequency comb (Toptica DFC CORE) is stabilized near $80$\,MHz and phase locked using the ULE-cavity-locked $689$\,nm laser as a reference.
The comb is used to stabilize the relative frequency and phase of all remaining excitation lasers with phase locks~\cite{Diddams2000}(via Toptica PFD and FALC 110 current servo).
This results in phase-synchronized optical frequencies for excitation that exhibit the stability of the reference cavity. 
The ULE reference cavity has a linear drift of $6$\,Hz/min that we compensate daily with spectroscopy on the $\gnd-\ex$ transition.
This drift is uncompensated during an experiment run time ($1$\,Hz repetition rate), though the resulting laser frequency shifts are highly correlated between all excitation lasers and remain much smaller than the excitation bandwidths.

The seed light is delivered to the injection-locked diodes over polarization-maintaining optical fibers. 
Fiber phase noise is actively suppressed to below $30$\,mrad root-mean-square. 
The paths for each laser contain individual acousto-optic modulators driven by ARTIQ-controlled Urukul DDS boards~\cite{Bourdeauducq2016} which provide both fast switching and independent phase control of the lasers and are used to create deterministic phase shifts in Ramsey experiments.

The laser parameters used in our demonstrations and information about atomic transitions specific to $^{88}$Sr are provided in Table~\ref{tab:beam_params}.
For the demonstrations in the main text, we use $\approx$~$10\,$mW-level optical powers and millimeter-scale $1/\text{e}^2$ intensity radii for the excitation lasers.
The laser polarization purity is enforced with polarizing beamsplitter cubes as the final optical element in each laser path to the atomic sample.

\begin{table}[ht]
\centering
\renewcommand{\arraystretch}{1.2}
\caption{Summary of laser parameters for the experiments demonstrated using ${}^{88}$Sr in the main text.}
\label{tab:beam_params}
\begin{tabular}{l  | c c c }
\hline\hline
Laser index & 1 & 2 & 3 \\
\hline
Coupled manifolds  & $\gnd-\ex$ & $^3P_1-{}^3S_1$ & ${}^3P_0-{}^3S_1$ \\
Transition label   &  $\ket g-\ket{s_\pm}$   & $\ket{s_\pm}-\ket v$ & $\ket e-\ket v$ \\
Transition linewidth $\gamma_i / 2\pi$ &$7.5$\,kHz&$3.9$\,MHz&$1.3$\,MHz\\
Wavelength (vac.) $\lambda_i$ (nm) & $689.449$ & $688.021$ & $679.289$ \\
Polarization  $\hat{\bm{\epsilon}}_i$ & $\hat{\mathbf{z}}$ & $\hat{\mathbf{z}} $ & $\hat{\mathbf{x}}$ \\
Delivery angle $|\theta_{3i}^\text{opt}|$ (deg) &  $59.64$  & $59.44$  & --- \\
$1/\textrm{e}^2$ radius $w_{i}$ (mm) & $0.5$ & $0.9$ & $0.9$ \\
Optical power $P_i$ (mW)\footnote{Narrow-line data from Sec~\ref{sec:dfres} uses variable optical powers.} & $18$\footnote{Simultaneous excitation method from Sec.~\ref{sec:method1}.\label{fn:1}}/$27$\footnote{Sequential excitation method from Sec.~\ref{sec:method2}.\label{fn:2}} & $11$/$13$ & $3.8$/$12$ \\
\hline\hline
\end{tabular}
\end{table}

\subsection{State readout}\label{app:readout}
After atomic state manipulation with the excitation lasers, we employ three types of state-selective imaging schemes for atomic state readout.
Each readout method uses a combination of push and repump pulses. 
Push pulses consist of $1$\,\textmu s pulses of light resonant with the $461$\,nm $\gnd-{}^1\!P_1$ transition 
with natural linewidth of $32$\,MHz that rapidly impart momentum to ground state atoms along the $\hat{\mathbf{x}}$-direction.
The repump pulses address the $679$\,nm $\clock-{}^3\!S_1$ and/or the $707$\,nm ${}^3\!P_2-{}^3\!S_1$ transitions and transfer population from $\clock$ and ${}^3\!P_2$ to the ground state via the ${}^3\!S_1$ and $\ex$ decay channels. 
Strategic combinations of these pulses lead to a mapping of atomic states to position.
After spatial separation of states, all atoms either decay or are pumped to the ground state and are imaged in fluorescence on a CMOS camera (Thorlabs CS505MU) with a $20$\,\textmu s $461$\,nm excitation pulse. 
The images are separated into ports based on cloud position and the population fraction is computed by the ratio of each port pixel count to the total pixel count.
A background image taken with no atoms in the imaging region is subtracted from all readout images.

To measure population in the $\gnd$ and $\ex$ states, a single push pulse is applied followed by a 1\,ms delay. In that delay time, the ground state atoms have moved $\approx700\,$\textmu m and are spatially separated from $\ex$ atoms. Additionally, the $\ex$ atoms have decayed to the ground state and the fluorescence imaging pulse interacts with all atoms equally.
There are two protocols for detection involving the clock state.
The first results in three spatially separated ports that contain (1) $\gnd$ atoms, (2) $\ex$ atoms, and (3) $\clock+{}^{3}\!P_2$ atoms. 
The second results in ports that contain (1) $\gnd$+$\ex$ atoms, (2) $\frac{3}{4}\,{}^{3}\!P_2$ atoms, and (3) $\clock$+$\frac{1}{4}\,{}^{3}\!P_2$ atoms. 
These protocols are described in more detail in~\cite{Carman2025}.
We measure the populations in each of the $\gnd$ and ${}^{3}\!P_J$ states by combining experimental runs with each of these imaging schemes.
Example readout images are provided in the inset of Fig.~\ref{fig:simulrabi} in the main text.

We verify that detection in each port achieves leakage and residual background counts below $1\%$ by preparing the entire atomic ensemble in each readout port and computing the population fraction, which confirms the ability to measure atomic state populations at the $1\%$ level.
Measurements of the $\ex$ population are underestimated by $4.5\%$ due to its spontaneous decay to the ground state during the finite push pulse duration.
Data presented in the main text are corrected for this effect by appropriate scaling of the atomic populations.

\subsection{Ramsey contrast decay analysis}\label{app:fits}
Here we describe the procedures for the fits used to quantify Ramsey contrast $1/\text{e}$ decay time for the DF and single-photon excitation methods shown in Fig.~\ref{fig:ramseydata}(c).
We parameterize the Ramsey contrast as a function of the Ramsey dark time $\tau$ using the fit function
\begin{equation}
    f(\tau; A,t_I,t_H) = A \exp\left[  -\left(\frac{\tau}{t_I}\right)^2 - \frac{\tau}{t_H} \right] \, ,
\end{equation}
where the $A$ is the free parameter accounting for contrast reduction due to imperfect $\pi$-pulses and $t_I$ and $t_H$ are the characteristic decay times for Gaussian (inhomogeneous) and exponential (homogeneous) decay processes, respectively.
We fit this function to the contrast decay data to extract the optimal fit parameters $\{A^*,t_I^*,t_H^*\}$ for each excitation scheme.
The quoted $1/\text{e}$ decay time is the intersection of the optimal fit $f(\tau;A^*,t_I^*,t_H^*)$ with the value $A^*/\text{e}$ while the uncertainty is from the intersection of the $90\%$ confidence band with the value $A^*/\text{e}$.

\section{Systematic considerations}\label{app:sys}

\subsection{Density shifts}

Interatomic collisions produce a frequency shift during interrogation that scales with the atomic density and \mbox{$s$-wave} scattering length $a$. For fermions this shift is strongly suppressed, as the Pauli exclusion principle forbids $s$-wave collisions. Typically, bosons in high-density optical lattice sites are limited by this density shift~\cite{Ludlow2015}. For our experiment, we benefit from the significantly lower density of a free-space sample, lowering the shifts relative to those seen in tightly confined atomic ensembles. 
For our $3\!\times\!10^6$ atoms, we find peak densities near $10^{12}\,\text{cm}^{-3}$, contributing shifts on the order of $50$\,Hz~\cite{Lisdat2009}. During free fall, ballistic expansion lowers our densities even further, bringing the shift down dramatically. We neglect this effect for the experiments and analysis in this work. Dilute, free-space ensembles of ${}^{87}$Sr could be used to further suppress this effect if required.

\subsection{Off-resonant scattering}\label{subsec:scat}
In both methods, off-resonant scattering via the short-lived $\ket v$ state, which has a decay rate $\Gamma_v = 72\times10^6\,\text{s}^{-1}$, poses limits on clock state excitation. 
A state $\ket{j}$ coupled to $\ket v$ with coupling strength $\Omega_{jv}$ and detuning $\Delta_{jv}$ scatters at a rate $R_{\text{sc},j}=\Gamma_v\Omega_{jv}^2/4\Delta_{jv}^2$. 
For the simultaneous method, lasers 1 and 2 additionally contribute to two-photon scattering from the ground state with rate $R_{\text{sc},g}=\Gamma_v\Omega_1^2\Omega_2^2/16\Delta_1^2\Delta_2^2$. 
We approximate the scattered population after time $T$ as the sum of the individual contributions from each state $j$
\begin{align}
    P_\text{sc}(T) \approx \sum_j \int_0^TR_{\text{sc},j}P_j(t) \mathrm{d} t \, ,
\end{align}
which is valid when all scattered populations remain small.

For method 1, population outside of $\ket g$ and $\ket e$ is only virtually populated, so our scattering during a pulse of duration $T$ will be dominated by $R_{\text{sc},g}$ and $R_{\text{sc},e}$, given by 
\begin{align}
    P_\text{sc}(T) = R_{\text{sc},g}\int_0^{T} P_g(t) \mathrm{d}t + R_{\text{sc},e}\int_0^{T} P_e(t) \mathrm{d} t.
\end{align}
For method 2, the two-photon pathway does not exist since lasers 1 and 2 are not on simultaneously.
However, single-photon scattering from laser 2 on $\ket {s_+}$ population must be considered, resulting in 
\begin{align}
    P_\text{sc}(T) = R_{\text{sc},s_+}\int_0^{T} P_{s_+}(t) \mathrm{d} t + R_{\text{sc},e}\int_0^{T} P_e(t) \mathrm{d} t.
\end{align}

We numerically evaluate these equations to estimate the total scattered population. 
We scale the results by a factor of $8/9$ to account for the portion of scattered population that returns to $\ket e$, set by the branching ratio $\beta_{ve}=1/9$. 
This leads to a slight disparity between the estimate of peak excitation fraction and coherent $\pi$-pulse fidelity.
We obtain excitation fraction reductions of $7.2\%$ and $2.9\%$ for methods 1 and 2, respectively.

\subsection{Velocity-dependent coupling strength}\label{app:sys_finite_detuning}
Here we quantify the limit of excitation fraction achievable for the two methods demonstrated due to finite detunings and finite Rabi frequencies. We assume perfect DF alignment of the excitation lasers.

\paragraph*{Method 1:}
The detuning of the excitation lasers is a function of atomic velocity $\mathbf{v}$ and, as a result, so is the three-photon Rabi frequency. Taking this effect into account, we modify Eq.~\eqref{eq:Rabi3v} in the main text by including velocity-dependent detuning terms $\Delta_1'(\mathbf{v}) = \Delta_1-\mathbf{k}_1\cdot\mathbf{v}$ and $\Delta_2'(\mathbf{v}) = \Delta_2-\mathbf{k}_3\cdot\mathbf{v}$ where we have made the substitution $\mathbf{k}_1+\mathbf{k}_2=\mathbf{k}_3$. The effective three-photon Rabi frequency is then
\begin{equation}\label{eq:Rabi3v_eff}
    \Omega_\text{3$\nu$}(\mathbf{v}) = \frac{\Omega_1 \Omega_2 \Omega_3}{4\Delta_2'(\mathbf{v})} \left(  \frac{1}{\Delta_1'(\mathbf{v})} - \frac{1}{\Delta_1'(\mathbf{v})+2\delta_B}\right) .
\end{equation}
We use detunings $\Delta_1,\Delta_2$ much larger than the Doppler width and compute the leading-order velocity-dependent effect on excitation fraction due to this inhomogeneity in Rabi frequency. 
Expanding Eq.~\eqref{eq:Rabi3v_eff} to first order in $\mathbf{k}_i\cdot\mathbf{v}$ so that $\Omega_{3\nu}(\mathbf{v}) \approx \Omega_{3\nu}(\mathbf{0})(1+\epsilon_{\mathbf{v}})$ we obtain
\begin{equation}
   \epsilon_{\mathbf{v}}= \left( \frac{2(\Delta_1+\delta_B)}{\Delta_1(\Delta_1+2\delta_B)}\mathbf{k}_1+
    \frac{1}{\Delta_2}\mathbf{k}_3\right)\cdot\mathbf{v}.
\end{equation}
We define $\mathbf{A}$ so that $\epsilon_\mathbf{v} = \mathbf{A}\cdot\mathbf{v}$.
It follows that $\langle \epsilon_\mathbf{v}^2 \rangle = |\mathbf{A}|^2\langle v_A^2 \rangle $, where $\langle v_A^2 \rangle = k_BT_A/m$ is the one-dimensional Maxwell-Boltzmann velocity variance for an ensemble of temperature $T_A$ in the direction of $\mathbf{A}$. 
Thus, we obtain
\begin{equation}
    \langle \epsilon_\mathbf{v}^2 \rangle = \frac{k_BT_A}{m}\left|  \frac{2(\Delta_1+\delta_B)}{\Delta_1(\Delta_1+2\delta_B)}\mathbf{k}_1+
    \frac{1}{\Delta_2}\mathbf{k}_3 \right|^2 .
\end{equation}

Upon driving the three-photon transition for a nominal $\pi$-pulse duration of $\pi/\Omega_{3\nu}(\mathbf{0})$, the velocity-dependent excitation fraction is $P_\text{exc}(\mathbf{v}) = (1+\cos(\pi\epsilon_\mathbf{v}))/2$.
Since $\epsilon_\mathbf{v}$ is a zero-mean Gaussian random variable, $\langle \cos(a\epsilon_\mathbf{v})\rangle = \exp(-a^2\langle \epsilon_\mathbf{v}^2\rangle/2)$.
It follows that the mean excitation fraction is
\begin{equation}
    \langle P_\text{exc}\rangle = \frac{1}{2}\left(1+\exp(-\pi^2\langle\epsilon_{\mathbf{v}}^2\rangle/2) \right).
\end{equation}
For our experimental parameters, including a temperature of 9\,\textmu K, we obtain a peak excitation fraction due to finite three-photon detunings that deviates from unity by less than $0.1\%$ and thus does not pose a limit in this work.

\paragraph*{Method 2:}
This method consists of resonant single-photon and Raman transitions. We use the generalized Rabi frequency as a function of velocity to estimate the peak excitation fraction from a thermal ensemble. For a Rabi frequency $\Omega_i$ and wavevector $\mathbf{k}_i$, an atom with velocity $\mathbf{v}$ is Doppler shifted from resonance by $\mathbf{k}_i\cdot\mathbf{v}$ and undergoes a Rabi oscillation with generalized Rabi frequency $\tilde{\Omega}_i = (\Omega^2_i+(\mathbf{k}_i\cdot\mathbf{v})^2)^{1/2}$. The peak excitation after a nominal $\pi$-pulse time of $\pi/\Omega_i$ is then $P_\text{exc,i} = \Omega_i^2/\tilde{\Omega}_i^2$. Expanding to lowest order in $\mathbf{k}_i\cdot\mathbf{v}/\Omega_i$, we have that $ P_\text{exc,i} \approx 1 - (\mathbf{k}_i\cdot\mathbf{v}/\Omega_i)^2$.
Then averaging over the thermal distribution of the atomic sample we obtain a mean excitation fraction of 
\begin{equation}
    \langle P_\text{exc,i} \rangle = 1 - \frac{k_i^2}{\Omega_i^2}\frac{k_BT}{m} .
\end{equation}
For the parameters in our demonstration, the finite Rabi frequency limits on the single-photon and Raman excitation infidelities are $<\!0.1\%$ and $0.2\%$, respectively, not significantly contributing to our overall excitation limits.

\subsection{Angular misalignment}\label{app:sys_angle}

For the excitation wavelengths in this work, the angles between the excitation lasers that satisfy the DF condition as shown in Fig.~\ref{fig:3v}(a) are $\theta_{31}^\text{opt} = 59.64^\circ$ and $\theta_{32}^\text{opt} = 59.44^\circ$. 
Any deviations from these optimal angles result in a residual uncanceled wavevector $\Delta\mathbf{k}=\mathbf{k}_1+\mathbf{k}_2-\mathbf{k}_3$ that in turn leads to velocity sensitivity in a Ramsey sequence.
Around optimal alignment, the residual wavevector contribution from misaligned lasers is well approximated by 
\begin{equation}
    \frac{|\Delta\mathbf{k}|}{|\mathbf{k}_1|} \approx \sqrt{\left(\delta\theta_{31}\right)^2+\left(\delta\theta_{32}\right)^2+\delta\theta_{32}\delta\theta_{31}} ,
\end{equation}
where $\delta\theta_{3i}$ is the deviation of laser $i$ from its optimal angle in radians.
Allocating all misalignment to the plane of excitation and splitting equally across lasers 1 and 2, our results suggest alignment inaccuracy of $<\!0.5\,$mrad per laser.
Misalignment out of the excitation plane can be lumped into a single excitation laser and was optimized by tuning the pitch of laser 3.
This was accomplished with Newport UTR46 rotation stages and NewFocus Picomotor Model 8302 
to point the excitation lasers at the atomic sample.

\subsection{Ensemble free-fall and free-space expansion}\label{app:sys_freefall}

As indicated in Fig.~\ref{fig:ramseydata}(c) in the main text, Ramsey contrast can
be limited in these demonstrations due to atomic ensemble free-fall out of the
finite excitation beam waists along with thermal expansion during the dark time.
In our demonstration the $1/\text{e}^2$ intensity radii $w_i$ for the three excitation
lasers range from 500 to 900\,\textmu m.

We work in a laboratory frame $\mathbf{r}=\{x,y,z\}$ with the origin at the common center of the excitation
beams. The displacement of an atom from
the axis of beam $i$ is the component of $\mathbf{r}$ transverse to $\hat{\mathbf{k}}_i$,
\begin{equation}
    |\mathbf{r}^\perp_i|^2 = |\mathbf{r}|^2 - (\mathbf{r}\cdot\hat{\mathbf{k}}_i)^2
    = r_i^2 + z^2.
    \label{eq:perp}
\end{equation}
Here $r_i \equiv \mathbf{r}\cdot(\hat{\mathbf{z}}\times\hat{\mathbf{k}}_i)$ is the in-plane distance perpendicular to the propagation direction of laser $i$, while the vertical
coordinate $z$ is shared between all three beams.

After the dark time $\tau$, free fall shifts the cloud center in the $\hat{\mathbf{z}}$-direction by
$\Delta z(\tau) = -g\tau^2/2$, where $g$ is the acceleration due to Earth's
gravity. 
The initial positions of the atoms in the ensemble are normally distributed.
The horizontal extent is radially isotropic with standard deviation
$\sigma_{r,0}$ and the vertical extent has standard
deviation $\sigma_{z,0}$. Ballistic expansion increases these to
\begin{equation}
    \sigma_j(\tau) = \sqrt{\sigma_{j,0}^2 + \frac{k_B T_j}{m}\tau^2},
    \label{eq:widths}
\end{equation}
where $T_j$ is the temperature in direction $j\in\{r,z\}$. In our results from Sec.~\ref{sec:ramsey} in the main text, the atomic
samples have temperatures of 9\,\textmu K (5\,\textmu K) in the $\hat{\mathbf{r}}$ ($\hat{\mathbf{z}}$) direction, corresponding to $1\sigma$ spatial extent of the cloud of $95$\,\textmu m ($145$\,\textmu m) after a typical $3$\,ms Ramsey dark time. These values are determined by time-of-flight imaging and are used for expansion modeling.
At time $\tau$ the single-atom position distribution is therefore the
three-dimensional Gaussian
\begin{equation}
\begin{aligned}
    f(\mathbf{r};\tau) =&
    \frac{1}{(2\pi)^{3/2}\,\sigma_r^2(\tau)\,\sigma_z(\tau)}\\
    \times&\exp\!\left(
        -\frac{x^2+y^2}{2\sigma_r^2(\tau)}
        -\frac{\big(z-\Delta z(\tau)\big)^2}{2\sigma_z^2(\tau)}
    \right).
    \label{eq:dist}
\end{aligned}
\end{equation}

The displacement of the atom from each beam center reduces the corresponding
single-photon Rabi frequency to
\begin{equation}
    \Omega_i'(\mathbf{r}) = \Omega_i^0 
    \exp\!\left(-\frac{r_i^2 + z^2}{w_i^2}\right).
    \label{eq:rabi1}
\end{equation}
We find the implied contrast reduction from the expectation value of the
maximum excitation fraction $\langle P_\text{exc}\rangle$ achievable starting
from an equal superposition state. The initial state-preparation pulse is assumed to be perfect since it occurs when the atoms have not fallen or expanded appreciably relative to the beams. 

For method 1, the effective coupling for three-photon clock excitation is proportional to the product of the three single-photon couplings, allowing us to write a reduction factor as
\begin{equation}
    \eta_{3\nu}(\mathbf{r}) \equiv \Omega_{3\nu}'(\mathbf{r})/\Omega_{3\nu}^{0}
    = \prod_{i=1}^{3}\exp\!\left(-\frac{r_i^2 + z^2}{w_i^2}\right),
    \label{eq:rabi3}
\end{equation}
where 
$\Omega_{3\nu}^{0}$ is the nominal three-photon Rabi frequency for an atom at the origin given by Eq.~\eqref{eq:Rabi3v} in the main text. 
A single atom at position $\mathbf{r}$ undergoes Rabi oscillations at the reduced rate $\Omega_{3\nu}'(\mathbf{r})$. 
The pulse area seen by the atom is set deterministically by its position in the beams.
We average over the position distribution $f(\mathbf{r};\tau)$ to compute the ensemble excitation fraction.

The closing $\pi/2$-pulse is tuned to the nominal three-photon
Rabi frequency $\Omega_{3\nu}^{0}$, i.e., its duration is fixed at
$T_{3\nu}^{\pi/2}=\pi/(2\Omega_{3\nu}^{0})$. An atom at $\mathbf{r}$ is therefore
under-rotated on the Bloch sphere and its excitation fraction is
\begin{equation}\label{eq:P_exc1_id}
    P_\text{exc,1}(\mathbf{r}) =
    \frac{1}{2}\left(1 + \sin\!\left(\frac{\pi}{2}\,\eta_{3\nu}(\mathbf r)\right)\right).
\end{equation}
Averaging over the ensemble
\begin{equation}\label{eq:Pexc_avg}
    \langle P_\text{exc,1}(\tau)\rangle =
    \int f(\mathbf{r};\tau)\,P_\text{exc,1}(\mathbf{r})\,\mathrm{d}^3r
    = \frac{1}{2}\left(1 + C(\tau)\right)
\end{equation}
identifies the Ramsey contrast as the expectation value 

\begin{equation}\label{eq:contrast}
\begin{aligned}
    C(\tau) &= \int f(\mathbf{r};\tau)\,
    \sin\!\left(\frac{\pi}{2}\eta_{3\nu}(\mathbf{r})\right)\, \mathrm{d}^3r.
\end{aligned}
\end{equation} 
Equation~\eqref{eq:contrast} has no closed form and we
evaluate it by numerical integration over $f(\mathbf{r};\tau)$, which yields a free-space contrast
decay limit of 9.4\,ms for the parameters used in our demonstration.

For method 2, closing the Ramsey sequence uses a Raman $\pi$-pulse followed by a single-photon $\pi/2$-pulse, set to nominal excitation times $T_R^\pi = \pi/\Omega_R^0$ and $T_1^{\pi/2} = \pi/(2\Omega_1^0)$. This gives an excitation fraction of 
\begin{equation}\label{eq:P_exc2}
\begin{aligned}
    P_{\textrm{exc},2}(\mathbf r) =  &\frac{1}{2}\bigg( \sin^2\left(\frac{\pi}{4}\eta_1(\mathbf r)\right) + \\
    &\sin^2\left(\frac{\pi}{2}\eta_R(\mathbf r)\right)\cos^2\left(\frac{\pi}{4}\eta_1(\mathbf r)\right) + \\
    &\sin\left(\frac{\pi}{2}\eta_1(\mathbf r)\right)\sin\left(\frac{\pi}{2}\eta_R(\mathbf r)\right) \cos \phi_1 \bigg), 
\end{aligned}
\end{equation}
where $\phi_1$ is the phase of the final applied $\pi/2$-pulse and we define reduction factors
\begin{align}
    \eta_R(\mathbf r)\equiv\Omega_R'(\mathbf r)/\Omega_R^0 = &\, \prod_{i=2}^{3}\exp\!\left(-\frac{r_i^2 + z^2}{w_i^2}\right)
\end{align}
and 
\begin{align}
    \eta_1(\mathbf r)\equiv\Omega_1'(\mathbf r)/\Omega_1^0 = &\, \exp\!\left(-\frac{r_1^2 + z^2}{w_1^2}\right).
\end{align}
This takes the form $P_\text{exc,2} = \frac{1}{2}\left(A + C(\tau) \cos \phi_1\right)$, allowing us to identify the contrast term as
\begin{equation}
    C(\tau) =  \int f(\mathbf{r};\tau)\,
    \sin\left( \frac{\pi}{2} \eta_1(\mathbf{r})\right)\sin\left( \frac{\pi}{2} \eta_R(\mathbf{r})\right)\,\mathrm{d}^3r,
\end{equation}
where we use the same ensemble spatial distribution from above.
Numerically evaluating this for the parameters in our demonstration results in a limit of 9.7\,ms. For both methods, this limit is close to but larger than our residual Doppler limit.

\subsection{Gaussian beam effects}\label{sec:wavefront}

The DF condition is perfectly satisfied for plane waves propagating at the optimal beam angles. However, for Gaussian beams the local wavevector is determined by a combination of propagation angle, wavefront curvature, and Gouy phase. These latter two contributions lead to deviations from the DF condition.
To determine the impact of this effect, we calculate the spatial phase due to a realistic Gaussian beam. 

The spatial phase of a Gaussian beam propagating in the $z$-direction with wavevector magnitude $k$ is given by
\begin{align}
    \Theta(\rho, z) = kz + \frac{k \rho^2}{2 R(z)} - \psi(z),
\end{align}
where $\rho$ is the radial position from the beam center, $z_R$ is the Rayleigh range, $R(z) = z\big(1+(z_R/z)^2\big)$ is the radius of curvature, and $\psi(z) = \arctan(z/z_R)$ is the Gouy phase.
The local wavevector for each beam is given by the gradient of its spatial phase $\mathbf{k}_i'=\nabla\Theta(\rho_i, z_i)$, which evaluates to
\begin{equation}
    \begin{aligned}
        \mathbf{k}_i'&(\rho_i,  z_i) 
        =\frac{k_i\rho_i}{R(z_i)}\hat{\bm{\rho}}_i +\\&\left ( k_i  +\frac{k_i\rho_i^2}{2} \frac{z_{R, i}^2-z_i^2}{(z_{R, i}^2+z_i^2)^2} -\frac{z_{R, i}}{z_{R, i}^2+z_i^2}   \right )\hat{\mathbf{z}}_i, 
    \end{aligned}
\end{equation}
where the spatial coordinates $\rho_i$ and $z_i$ are defined in the frame of laser $i$ through the center of the atomic cloud. Since the length scale of the atomic cloud is four orders of magnitude smaller than the Rayleigh range, we simplify to
\begin{align}
    \mathbf{k}'_{i}(\rho_i,z_i)  \approx  \frac{k_i\rho_iz_i}{z^2_{R,i}}\hat{\bm{\rho}}_i+ \bigg(k_i+\frac{k_i\rho_i^2}{2z^2_{R,i}}-\frac{1}{z_{R,i}}\bigg)\hat{\mathbf{z}}_i .
\end{align}
The nominal wavevector is $k_i\hat{\mathbf{z}}_i$ and all other terms comprise the Gaussian beam correction $\delta \mathbf{k}_i$. 
We compute the residual wavevector due to these terms $\delta \mathbf{k} = \delta\mathbf{k}_1+\delta\mathbf{k}_2-\delta\mathbf{k}_3$ for our laser geometry and sizes and find a maximum magnitude of 0.6\,m$^{-1}$ over the spatial extent of the cloud, corresponding to $|\delta\mathbf{k}|/|\mathbf{k}_1| = 7\times10^{-8}$.
Applying this maximum value to all atoms in the ensemble gives an upper bound on the residual Doppler width due to Gaussian beam propagation of 7\,mHz for a 9\,\textmu K sample.

\subsection{Peak excitation infidelity budget}\label{app:fid}
Here we summarize the dominant contributions to the peak excitation fraction infidelity in our system.
We include the effects detailed in the preceding sections along with effects modeled in Appendix~\ref{app:sim_model} below such as finite laser size, ensemble temperature, and transient state population.
Table~\ref{tab:fid} shows these contributions for both methods presented in the main text.

\begin{table}[ht]
\centering
\renewcommand{\arraystretch}{1.2}
\caption{Summary of dominant contributions to peak excitation infidelity for a simultaneous (method 1) and sequential (method 2) DF $\pi$-pulse given the excitation parameters from Sec.~\ref{sec:method1} and Sec.~\ref{sec:method2}, respectively.}
\label{tab:fid}
\begin{tabular}{l | c c | c  }
\hline\hline
Mechanism & Method 1 & Method 2\footnote{Quantities determined via simulation for method 2 use the same model described in Appendix~\ref{app:sim_model} but with additional time-varying envelopes on the individual couplings to account for the sequential pulses.} & Details \\
\hline
Transient state population & $0.4\%$ &   $<\!0.1\%$& App.~\ref{app:sim_model} \\
Intensity inhomogeneity       & $14.6(7)\%$ & $2.5(5)\%$  & App.~\ref{app:sim_model}  \\
Off-resonant scattering\footnote{The off-resonant scattering reported does not account for the small portion of scattering events that return the atom to the $\ket e$ state via (1) the $\beta_{ve}=1/9$ branching ratio from $\ket v$ and (2) the $<\!1\%$ scattered population that returns to the ground state and re-excites to $\ket e$. This population represents incoherent excitation and contributes to the peak excitation fraction, but not to the coherent $\pi$-pulse fidelity.}  
& $7.2\%$ & $2.9\%$ & App.~\ref{subsec:scat}  \\
Velocity-dependent coupling   & $<\!0.1\%$ &  $0.2\%$ & App.~\ref{app:sys_finite_detuning} \\
Intermediate-state decay         & --- & $1.4\%$  & Sec.~\ref{subsec:ex2} \\
Residual Doppler              & $<\!0.1\%$ &  $<\!0.1\%$ & App.~\ref{app:sys_angle}\\
\hline
\textbf{Fidelity (modeled)}\footnote{The modeled fidelity is determined via full simulation to account for interactions between the effects, rather than computed as a product of individual mechanisms.}  & $79.5(6)\%$ &  $93.0(5)\%$  &    \\
\textbf{Fidelity (measured)} & $76(2)\%$ & $91(1)\%$  & Fig.~\ref{fig:simulrabi}/\ref{fig:seq}      \\
\hline\hline
\end{tabular}
\end{table}

\section{Three-photon excitation model}\label{app:sim_model}

The simulations described here model the full coherent and dissipative dynamics of the three-photon simultaneous excitation method from Sec.~\ref{sec:method1} in the main text. 
We describe the framework for our numerical simulations that we compare against our measured state populations, such as in Fig.~\ref{fig:simulrabi}. 
While the analytic Rabi frequency from Eq.~\eqref{eq:Rabi3v} captures the idealized three-photon coupling, making a direct comparison with experiment additionally requires modeling transient population of the intermediate states, off-resonant scattering, finite pulse-edge transients, AC Stark shifts, and the inhomogeneous response of a finite-temperature ensemble in finite-sized beams. 

We model the system with six relevant energy levels and three optical laser fields for excitation.
We consider only the energy levels that are directly coupled by our excitation lasers, namely states $\ket g\!,\, \ket{s_-}\!,\, \ket{s_+}\!,\,\ket v$, and $\ket e$, as defined in the main text, together with the metastable $\ket d$ manifold that serves as the sixth level without differentiating between its Zeeman levels. 
Re-excitation from this dark state is not possible (as no lasers couple this manifold to any other) and it acts solely as an incoherent decay channel.
States outside the modeled pathway can in principle be populated via $\ket{s_0}$. However, population in $\ket{s_0}$ is strongly suppressed as it can only be reached via residual polarization impurity of laser 1 weakly coupling $\ket g$ and $\ket{s_0}$ and spontaneous decay from $\ket{v}$, which branches into all three $\ket s$ sublevels. Population that does accumulate in $\ket{s_0}$ can then be off-resonantly coupled by laser 2 to the unmodeled ${}^3S_1, \mj{\pm1}$ states. Each step in this chain is strongly suppressed by ensuring high polarization purity and sufficient intermediate-state detunings $\Delta_1+\delta_B$ and $\Delta_2$. The resulting leakage is therefore a higher-order effect and does not alter the simulated populations to within our tolerances. We omit these states from the reduced model.

The time-dependent solutions to the Schr\"odinger equation $i\hbar\partial_t\ket{\Psi_k(t)} = H_a\ket{\Psi_k(t)}$ for the atomic Hamiltonian $H_a$ are of the form $\ket{\Psi_k(t)} = \ket{k}e^{-i\omega_kt}$, where $\hbar\omega_k$ are the energies of each state with the ground state energy set to $0$.
The electric dipole interaction from the excitation lasers leads to a Hamiltonian of the form $H_\text{int} = -\mathbf{d}\cdot\mathbf{E}$, where $\mathbf{d}$ is the electric-dipole operator and $\mathbf{E}$ is the superposition of all laser fields used for excitation. 
With each laser having real electric field amplitude $E_i$ and polarization $\hat{\bm{\epsilon}}_i$, we have
\begin{align}
    \mathbf{E}(\mathbf{r},t) = \frac{1}{2}\sum_{i=1}^3E_i \hat{\bm{\epsilon}}_i\left[ e^{-i(\mathbf{k}_i\cdot \mathbf{r}-\omega_i^\ell t)} + e^{i(\mathbf{k}_i\cdot \mathbf{r}-\omega_i^\ell t)} \right].
\end{align}
The total Hamiltonian governing state evolution is then $H = H_a + H_\text{int}$.

We use a nested rotating frame so that each manifold rotates at the sum of preceding excitation laser frequencies. 
This unitary transformation is $U(t)=e^{-iGt}$ with generator

\begin{equation}
\begin{aligned}
    G &= \omega_1^\ell\big(\ket{s_+}\bra{s_+} +   \ket{s_-}\bra{s_-} \big)\\
    &\quad+(\omega_1^\ell+\omega_2^\ell)\ket{v}\bra{v} +(\omega_1^\ell+\omega_2^\ell-\omega_3^\ell)\ket{e}\bra{e} ,
\end{aligned}
\end{equation}
where $\omega_i^\ell$ are the optical frequencies of the excitation lasers as defined in the main text. Because laser 1 is polarized orthogonal to $\mathbf{B}$, it drives both $\ket g -\ket{s_+}$ and $\ket g - \ket{s_-}$ couplings, rotating at the same frequency $\omega_1^\ell$. The Hamiltonian transforms as $\tilde H = UHU^\dagger +i\hbar(\partial_tU)U^\dagger$. After discarding fast, counter-rotating terms (the rotating-wave approximation), we have a time-independent atomic Hamiltonian, which can be written as
\begin{equation}
\begin{aligned}
    \tilde H_\text{a}/\hbar =& -(\Delta_1+2\delta_B) \ket{s_{-}}\bra{s_{-}}-\Delta_1 \ket{s_{+}}\bra{s_{+}}\\ 
    &- \Delta_2 \ket{v}\bra{v} 
    - \Delta_3\ket{e}\bra{e}.
\end{aligned}
\end{equation}
For an atom moving at velocity $\mathbf{v}$ in the lab frame, the detunings transform as $\Delta_1 \rightarrow \Delta_1 - \mathbf{k}_1\cdot \mathbf{v}$,  $\Delta_2\rightarrow \Delta_2 - (\mathbf{k}_1+\mathbf{k}_2)\cdot \mathbf{v}$, and $\Delta_3\rightarrow \Delta_3 - (\mathbf{k}_1+\mathbf{k}_2-\mathbf{k}_3)\cdot \mathbf{v} = \Delta_3 - \Delta \mathbf{k}\cdot \mathbf{v}$. This is the only velocity-dependent element of our model.

The interaction term, manifesting as our remaining off-diagonal elements, can be written as
\begin{equation}
    \tilde H_\text{int}/\hbar = \sum_{k\neq l} \frac{ \Omega_{kl}}{2} \ket{k}\bra{l} , 
\end{equation}
where $\Omega_{kl}$ is the Rabi frequency of the $\ket{k}-\ket{l}$ transition from the excitation laser field $\mathbf{E}$ taking into account the appropriate polarization couplings and Clebsch-Gordan coefficients.
The experimental values of $\Omega_{kl}$ are defined in the main text as $\Omega_1 = \Omega_{gs_+}=\Omega_{gs_-}$, $\Omega_2 = \Omega_{s_+v}= \Omega_{s_-v}$ and $\Omega_3 = \Omega_{ve}$. The Clebsch-Gordan coefficients for the $\ket g -\ket{s_+}$ and $\ket g -\ket{s_-}$ pathways enter with opposite signs and are the source of the destructive interference in Eq.~\eqref{eq:Rabi3v}. 
We propagate the density matrix $\rho$ in this rotating frame using the Lindblad master equation
\begin{equation}
\dot \rho=-\frac{i}{\hbar}[\tilde H,{\rho}]+ \sum_{k\neq l}\left(A_{kl}{\rho}A_{kl}^{\dagger}-\frac{1}{2}\{A_{kl}^{\dagger}A_{kl}, \rho  \} \right) ,
\label{eq:Lindblad_ME}
\end{equation}
with $\tilde H = \tilde H_a + \tilde H_\text{int}$ and collapse operators $A_{kl}=\sqrt{\gamma_{kl}}\ket{l}\bra{k}$ describing spontaneous emission from state $\ket{k}$ to $\ket{l}$ at partial decay rate $\gamma_{kl}$. We initialize the system with $\rho(0) = \ket g \bra g$.
For given initial conditions and laser parameters, Eq.~\eqref{eq:Lindblad_ME} is solved numerically using \textsc{QuTiP} \cite{Johansson2013}.

\begin{figure}[tbp]
    \centering
    \includegraphics[width=\linewidth]{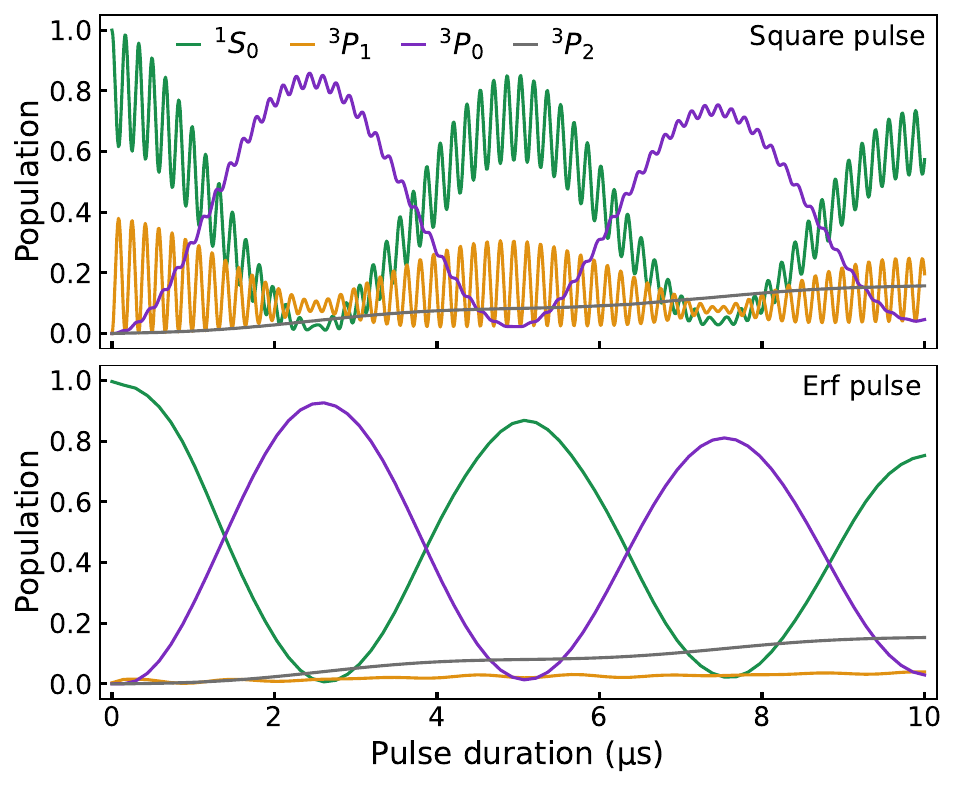}
    \caption{Simulated population dynamics for a single stationary atom undergoing simultaneous three-photon excitation using square (top) and error-function ramped (bottom) laser pulses.
    Normalized state populations for each state are plotted versus pulse duration.
    Use of an error-function-shaped ramp for the excitation pulses leads to the suppression of off-resonant Rabi oscillations from laser 1 compared to square pulses.}\label{fig:app}
\end{figure}

We operate in the small-detuning regime for laser 1 (i.e., $\Delta_1 \approx\Omega_1$) to increase the three-photon Rabi frequency at fixed laser power.
In this regime, a square-pulse drive contains high-frequency Fourier components that would off-resonantly drive population into the intermediate $\ket{s_\pm}$ states.
These transient effects can be suppressed by ramping on our excitation beams with rise times slower than the time scale set by the inverse detunings, dominated by $1/\Delta_1$. 
This is modeled with a time-dependent envelope $u(t)$ that modifies the excitation Rabi frequency given by
\begin{equation}
    u(t) =  \frac{1}{4}\left (   1 + \text{erf}\left(\frac{t}{\tau_\text{rise}}\right)\right )\left (   1 -\text{erf}\left(\frac{t-T}{\tau_\text{fall}}\right)\right )
\end{equation}
with time constant $\tau_\text{rise/fall} \approx 100\,\text{ns}$ and pulse durations $T$.
In Fig.~\ref{fig:app} we compare the state evolution for a stationary single atom evolving under Eq.~\eqref{eq:Lindblad_ME} with square laser pulses to that of ramped pulses to demonstrate the impact. 
The bottom panel shows the resulting Rabi oscillations at identical simulation parameters to those of Sec.~\ref{subsec:ex1} of the main text.

In our simulations, we include the finite temperature and spatial extent of the atomic ensemble via Monte Carlo sampling of the position and velocity distributions and ensemble averaging. Because the average displacement and any residual recoil of an atom during the pulse sequence are small compared to the beam size, we treat each atom's position and velocity as fixed over the excitation period, and assign it a randomly sampled $(\mathbf{r}, \mathbf{v})$ per shot.
We assume the velocities $v_j$ of the atoms in the thermal cloud are Maxwell-Boltzmann distributed and the positions $r_j$ are independently Gaussian distributed in each direction $\hat{\mathbf{j}}$, giving sampling distributions

\begin{align}
f_v(\mathbf{v})&= \prod_{j=x,y,z}\frac{1}{\sqrt{2\pi k_B T_j/m}} e^{-\frac{v_j^2}{2k_B T_j/m}}\label{eq:Pv}\\
f_r(\mathbf{r})&=\prod_{j=x,y,z}\frac{1}{\sqrt{2\pi\sigma_{r,j}^2}} e^{-\frac{r_j^2}{2\sigma_{r,j}^2}}\label{eq:Px}
\end{align}
with sample temperature $T_j$ and spatial extent $\sigma_{r, j}$ in each direction.

The effect of finite spatial extent of the atomic cloud relative to the beam size manifests as non-uniform coupling of the drive lasers (Appendix~\ref{app:prep} and Table~\ref{tab:beam_params}). For a given atom position, the Rabi frequency from each laser $i$ is reduced to 
\begin{equation}\label{eq:rabi_space}
\Omega_{i}'=\Omega_{i}^0e^{-| \mathbf r - (\mathbf r \cdot \hat{\mathbf k}_i)\hat{\mathbf k}_i |^2/w_{i}^{2}},
\end{equation}
where $| \mathbf r - (\mathbf r \cdot \hat{\mathbf k}_i)\hat{\mathbf k}_i |$ is the distance from the atom to the propagation axis of the laser.

The simulation results shown in Fig.~\ref{fig:simulrabi} in the main text are obtained as follows. We take $N$ random samples for $\mathbf{v}=\{v_{x},v_y, v_z\}$ and $\mathbf r = \{x,y,z\}$ from the distributions given by Eq.~\eqref{eq:Pv} and Eq.~\eqref{eq:Px}, respectively. 
For the $n$th sample, we calculate $\Omega_{i}'$ and $\mathbf{k}_{i}\cdot\mathbf{v}$ for all lasers. 
We solve Eq.~\eqref{eq:Lindblad_ME} numerically over a desired evolution time $t_{f}$ to calculate the final density matrix ${\rho}_{n}(t_{f})$. 
The population of each state $\ket{k}$ is then computed as $P_{n}(\ket{k})=\text{Tr}[{\rho}_{n}(t_{f})\ket{k}\bra{k}]$. 
The numerical data points in the figures are the mean values from all $N$ samples $\mathbb E[P(\ket{k})]=\frac{1}{N}\sum_{n=1}^{N}P_{n}(\ket{k})$. 
The sample size is chosen to be $N=1000$, large enough that all Monte Carlo sampling errors on the population estimates converge to $<\!0.1\%$.

\section{Sequential excitation model}\label{app:3vseq}

The utility of the sequential excitation method for phase-sensitive sensor protocols depends on the velocity-dependent phase evolution of the ground and clock state coherence during the excitation pulses themselves and during any subsequent dark time. 
First, we will reduce the four-level system of Fig.~\ref{fig:3v} to an effective three-level Hamiltonian in a unified rotating frame.
We will then construct the relevant propagators for analyzing arbitrary sequences of pulses and dark times. 
Finally, we will derive an expression for the phase readout of a standard Ramsey interferometer sequence.

\subsection{Effective three-level system}

The full level structure of Fig.~\ref{fig:3v} contains four primary states relevant to the sequential excitation. The Raman virtual state $\ket v$ is far-detuned from any single-photon resonance, with $\Delta_2 = 2\pi \times 400\,\text{MHz}$ much larger than the Raman drive amplitudes and the Doppler spread. The off-resonant scattering and accumulated population via the ${}^3\!S_1$ pathway over the sub-microsecond sequence of pulses are negligible. 
These conditions satisfy our adiabatic elimination of ${}^3\!S_1$ and result in a two-photon coupling between $\ket {s_+}$ and $\ket e$ with Rabi frequency $\Omega_R$ and wavevector $\mathbf{k}_R$, as defined in the main text.  
The remaining states form a three-level system with states $\ket g,\,\ket{s_+}$, and $\ket e$. 

The sequential excitation process consists of laser 1 driving the $\ket{g}-\ket{s_+}$ transition and a subsequent Raman coupling with lasers 2 and 3 driving the $\ket{s_+}-\ket e$ transition. 
We use a single unified rotating frame defined via the unitary $U=e^{-iGt}$ with generator
\begin{align}
    \label{eq:RWA}
    G = \omega_1^\ell \ket{s_+} \bra{s_+} + (\omega_1^\ell+\omega_2^\ell-\omega_3^\ell)\ket{e}\bra e  .
\end{align}
We make the Hamiltonian transformation $\tilde{H} = UHU^\dagger + i\hbar (\partial_t U)U^\dagger$ to construct time-independent Hamiltonians for both excitation pulses and free evolution.
The single-photon and Raman transition detunings are $\delta_1 = \omega_1^\ell - \omega_1$ and $\delta_R = (\omega_2^\ell-\omega_3^\ell) - (\omega_2-\omega_3)$, as defined in the main text.
The Doppler effect modifies these detunings by $\delta_i\to \delta_i - \mathbf{k}_i\!\cdot\!\mathbf{v}$. 
Under the DF condition ($\mathbf{k}_R = -\mathbf{k}_1$), the Doppler shifts are equal and opposite.

The relative phase evolution between the three states for any sequence of pulses is built from three possible operations: free evolution between pulses, the single-photon drive, and the two-photon Raman drive. 
We assume square-edge pulses for this analysis as a simplified model that captures the DF behavior.
In the rotating frame established above, we can determine the unitary propagator corresponding to each of these operations for $\pi$- and $\pi/2$-pulses in the three-level Hilbert space, and then apply them to the excitation and Ramsey schemes from the main text.

\subsection{Time-evolution operators}
In the rotating frame defined by Eq.~\eqref{eq:RWA} with basis $\{\ket g, \ket{s_+}, \ket e\}$, the Hamiltonians for free evolution, single-photon driving, and Raman driving are 
\begin{align}
    \tilde{H}_f/\hbar = &\begin{pmatrix}
        0 & 0 & 0\\
        0 &-\delta_1 &0\\
        0 & 0 & -(\delta_1 + \delta_R)
    \end{pmatrix}, \\[1ex] 
    \tilde{H}_1/\hbar = &\begin{pmatrix}
        0 & \Omega_1/2 & 0\\
        \Omega_1/2 &-\delta_1 &0\\
        0 & 0 & -(\delta_1 + \delta_R)
    \end{pmatrix},\\
    \intertext{and}
    \tilde{H}_R/\hbar = &\begin{pmatrix}
        0 & 0 & 0\\
        0 &-\delta_1 &\Omega_R/2\\
        0 & \Omega_R/2 & -(\delta_1 + \delta_R)
    \end{pmatrix} ,
\end{align}
respectively.
The time-evolution operators (propagators) are given by $U_x(T) = \exp(-i \tilde{H}_xT/\hbar)$ with $x\in \{f,1,R\}$.
A general state of the three-level Hilbert space, written as $\ket{\psi(t)} = \sum_k c_k(t) \ket{k} = c_g(t)\ket{g}+c_{s}(t)\ket{s_+}+c_e(t)\ket{e}$, evolves via $\ket{\psi(t+T)} = U_x(T)\ket{\psi(t)}$ for each of the propagators.

\subsubsection{Free evolution}
The time-evolution operator for free evolution under the Hamiltonian $\tilde H_f$ is given by
\begin{align}
    U_f(T) = \begin{pmatrix}
        1 & 0 & 0\\
        0 & e^{i\delta_1 T} & 0 \\
        0 & 0 & e^{i(\delta_1+\delta_R)T}
    \end{pmatrix},
\end{align}
where $T$ is the duration of the free evolution.
After this evolution, each state maintains its relative populations $|c_k|^2$ but accumulates relative phases.
The intermediate state $\ket{s_+}$ and clock state $\ket{e}$ accrue phase relative to $\ket{g}$ at a rate given by $\delta_1$ and $\delta_1+\delta_R$, respectively.

\subsubsection{Single-photon excitation}
During a single-photon excitation pulse, the clock state $\ket{e}$ is decoupled and evolves only via its diagonal phase term, while the upper block of the Hamiltonian $\tilde H_1$ induces standard two-level Rabi oscillations. 
The propagator is then a direct sum over the two blocks
\begin{align}
    U_1(T) = U_{gs_+}(T) \oplus e^{i(\delta_1+ \delta_R)T}.
\end{align}
The first block $U_{gs_+}$ is the propagator for a two-level system driven near resonance~\cite{steck}
\begin{equation}
\begin{aligned}
U_{gs_+}&(T) = e^{i\delta_1 T/2} \\
&\times \begin{pmatrix}
\cT\!-\!i\tfrac{\delta_1}{\tilde\Omega_1}\sT &\!\! -i\tfrac{\Omega_1}{\tilde\Omega_1}\sT \\
-i\tfrac{\Omega_1}{\tilde\Omega_1}\sT & \!\! \cT\!+\!i\tfrac{\delta_1}{\tilde\Omega_1}\sT
\end{pmatrix},
\end{aligned}
\end{equation}
where $\theta_1 = \tilde \Omega_1 T$ with generalized Rabi frequency $\tilde \Omega_1^2 = \Omega_1^2+\delta_1^2$.
We consider the special cases of $\pi/2$- and $\pi$-pulses, and compute $U_{gs_+}$ to leading order in $\delta_1/\Omega_1$. 
For a $\pi/2$-pulse, $T_1^{\pi/2} = \pi/(2\Omega_1)$ and we obtain
\begin{equation}
\begin{aligned}
 U_{gs_+}(T_1^{\pi/2}) & \approx \frac{e^{i\delta_1 T_1^{\pi/2}/2}}{\sqrt{2}}
\begin{pmatrix}
1 - i\delta_1/\Omega_1 & -i \\
-i & 1 + i\delta_1/\Omega_1
\end{pmatrix}.
\end{aligned}
\end{equation}
The $\pm i\delta_1/\Omega_1$ corrections on the diagonal elements encode the Bloch-vector tilt error produced by a detuned $\pi/2$ rotation. 
The off-diagonal couplings are unaffected to leading order. 
For a $\pi$-pulse, $T_1^{\pi} = \pi/\Omega_1$ and we obtain
\begin{equation}
\begin{aligned}
U_{gs_+}(T_1^{\pi})
&\approx -ie^{i\delta_1 T_1^\pi/2}
\begin{pmatrix}
\delta_1/\Omega_1 & 1 \\
1 & -\delta_1/\Omega_1
\end{pmatrix}.
\end{aligned}
\end{equation}
The diagonal elements correspond to incomplete population transfer, while the off-diagonal elements correspond to state inversion and upper block phase accumulation.

\begin{table*}[t]
\centering
\caption{Propagators for the four pulse cases used in our sequential excitation protocol. Each block acts on the relevant two-level subspace ($\{\ket g, \ket{s_+}\}$ for single-photon, $\{\ket{s_+}, \ket e\}$ for Raman). The decoupled state in each case accrues a diagonal phase as listed. All expressions are valid to first order in $\delta_i/\Omega_i$.}
\label{tab:propagators}
\renewcommand{\arraystretch}{1.5}
\setlength{\tabcolsep}{10pt}
\begin{tabular}{l | c c c}
\hline\hline
Pulse & Pulse-area condition & Active-block propagator & Decoupled-state evolution \\
\hline
Single-photon $\pi/2$ &
$T = \pi/(2\Omega_1)$ &
$\dfrac{e^{i\delta_1 T/2}}{\sqrt 2}
  \begin{pmatrix} 1 - i\delta_1/\Omega_1 & -i \\[2pt] -i & 1 + i\delta_1/\Omega_1 \end{pmatrix}$ &
\multirow{2}{*}[-3ex]{$e^{i(\delta_1+\delta_R)T}$ on $\ket e$} \\[5ex]

Single-photon $\pi$ &
$T = \pi/\Omega_1$ &
$-i\,e^{i\delta_1 T/2}
  \begin{pmatrix} \delta_1/\Omega_1 & 1 \\[2pt] 1 & -\delta_1/\Omega_1 \end{pmatrix}$ &
\\

\hline

Raman $\pi/2$ &
$T = \pi/(2\Omega_R)$ &
$\dfrac{e^{i(\delta_1+\delta_R/2)T}}{\sqrt 2}
  \begin{pmatrix} 1 - i\delta_R/\Omega_R & -i \\[2pt] -i & 1 + i\delta_R/\Omega_R \end{pmatrix}$ &
\multirow{2}{*}[-3ex]{$1$ on $\ket g$}\\[5ex]

Raman $\pi$ &
$T = \pi/\Omega_R$ &
$-i\,e^{i(\delta_1+\delta_R/2)T}
  \begin{pmatrix} \delta_R/\Omega_R & 1 \\[2pt] 1 & -\delta_R/\Omega_R \end{pmatrix}$ &
\\
\hline\hline
\end{tabular}
\end{table*}

\subsubsection{Raman excitation}

Similar to the single-photon case above, here $\ket g$ is decoupled and the propagator can be written as
\begin{align}
    U_R(T) = 1 \oplus U_{s_+e}(T)  ,
\end{align}
where the lower block $U_{s_+e}$ is another effective two-level system with a near-resonant drive.
In this case, there are two nonzero diagonal elements and the trace contributes a common diagonal shift $-(\delta_1 + \delta_R/2)$, which produces a phase $e^{i(\delta_1 + \delta_R/2)T}$ acting on both $\ket{s_+}$ and $\ket e$ during the pulse. The remaining block is again a standard driven two-level system, now with effective detuning $\delta_R$ and generalized Rabi frequency $\tilde\Omega_R^2 = \Omega_R^2 + \delta_R^2$, resulting in a drive angle $\theta_R = \tilde \Omega_R T$. Combining both contributions,
\begin{equation}
\begin{aligned}
U_{s_+e}&(T) = e^{i(\delta_1+\delta_R/2)T} \\
&\times \begin{pmatrix}
\cTR\!-\!i\tfrac{\delta_R}{\tilde\Omega_R}\sTR & \!\!\! -i\tfrac{\Omega_R}{\tilde\Omega_R}\sTR \\
-i\tfrac{\Omega_R}{\tilde\Omega_R}\sTR & \!\!\! \cTR\!+\!i\tfrac{\delta_R}{\tilde\Omega_R}\sTR
\end{pmatrix}.
\end{aligned}
\end{equation}
The common-diagonal prefactor $e^{i(\delta_1+\delta_R/2)T}$ is present on every matrix element of the lower block, regardless of pulse area, and is the origin of the velocity-dependent phase imprinted by the Raman pulse.
Physically, during the pulse the population spends roughly half of $T$ in $\ket{s_+}$ precessing at rate $\delta_1$, and half in $\ket e$ precessing at rate $\delta_1 + \delta_R$, giving an average rate of phase accumulation of $\delta_1 + \delta_R/2$ relative to the ground state $\ket{g}$.

Specializing to the pulse areas used in our protocol, the Raman $\pi/2$-pulse has duration $T_R^{\pi/2} = \pi/(2\Omega_R)$, resulting in
\begin{align}
U_{s_+e}(T_R^{\pi/2}) &\approx \frac{e^{i(\delta_1+\delta_R/2)T_R^{\pi/2}}}{\sqrt{2}} \notag \\
\times{}&\begin{pmatrix}
1 - i\delta_R/\Omega_R & -i \\
-i & 1 + i\delta_R/\Omega_R
\end{pmatrix}, \\
\intertext{and the Raman $\pi$-pulse ($T_R^\pi = \pi/\Omega_R$) gives}
U_{s_+e}(T_R^\pi) &\approx -ie^{i(\delta_1+\delta_R/2)T_R^\pi} \notag \\
\times{}&\begin{pmatrix}
\delta_R/\Omega_R & 1 \\
1 & -\delta_R/\Omega_R
\end{pmatrix} \, ,
\end{align}
to leading order in $\delta_R/\Omega_R$.
The structure mirrors the single-photon case. 
Each $\pi/2$-pulse imprints a Bloch-vector tilt of order $\delta_R/\Omega_R$ on the diagonal entries, while each $\pi$-pulse leaves a corresponding fractional population leakage of order $\delta_R/\Omega_R$. 
The distinction between Raman and single-photon pulses is the global prefactor $e^{i(\delta_1+\delta_R/2)T}$, which carries an additional $\delta_1$-dependent phase contribution absent in the single-photon case, a consequence of the $\ket{s_+}-\ket e$ block sharing the diagonal entry $-\delta_1$. 
A summary of these computed propagators can be found in Table~\ref{tab:propagators}.

\subsection{Phase dynamics}

Using the propagators above, we can extract the relative phase between two states at any point in a pulse sequence. 
The phase $\phi_{ab}$ between two states $\ket{a}$ and $\ket{b}$ after a pulse sequence is computed with the argument of the ratio of the state coefficients
$\text{arg}\left(c_a/c_b\right)$ after applying the corresponding unitaries.

\subsubsection{Phase imprint of a sequential opening pulse}

We now compute the detuning-dependent phase imprinted on the $\ket g-\ket e$ coherence by an opening sequential excitation starting from the clock state. 
The sequence consists of a Raman $\pi/2$-pulse, an inter-pulse free-evolution with duration $T_s$, a single-photon $\pi$-pulse, and a subsequent dark time $\tau$. The total propagator is
\begin{align}
    U_\text{open}(\tau) = U_f(\tau)\,U_{1}(T_1^{\pi})\,U_f(T_s)\,U_{R}(T_R^{\pi/2}).
\end{align}
The opening Raman $\pi/2$-pulse takes the system to a superposition of $\ket e$ and $\ket {s_+}$. 
This resulting state is given by 
\begin{align}
     \begin{pmatrix} c_g \\ c_s \\ c_e \end{pmatrix} = \begin{pmatrix} 0 \\ 0 \\ 1 \end{pmatrix} \xrightarrow{\,U_R^{\pi/2}\,}
    \frac{e^{i(\delta_1 + \delta_R/2) T_R^{\pi/2}}}{\sqrt 2}
    \begin{pmatrix} 0 \\ -i \\ e^{i \delta_R/\Omega_R} \end{pmatrix}.
\end{align}
The first gap of duration $T_s$ adds the corresponding diagonal phases to $c_s$ and $c_e$. The single-photon $\pi$-pulse then transfers the $\ket{s_+}$ amplitude to $\ket g$ via the off-diagonal entry of $U_1^{\pi}$, picking up the common-diagonal prefactor $-ie^{i\delta_1 T_1^\pi/2}$.  The subsequent dark time $\tau$ again adds the corresponding phases to $c_s$ and $c_e$. Composing these operations, the matrix elements of interest are
\begin{align}
    \bra{g}U_\text{open}(\tau)\ket{e} &= c_g(\tau) \\
    \bra{e}U_\text{open}(\tau)\ket{e} &=  c_e(\tau).
\end{align}
The relative phase between $\ket g$ and $\ket e$ at the end of the sequence, $\phi_{ge}^\text{open}(\tau) = \arg(c_e(\tau)/c_g(\tau))$, is
\begin{equation}
\begin{aligned}
    \phi_{ge}^\text{open}(\tau) =& (\delta_1 + \delta_R)\tau + \phi_\text{fp},
\label{eq:phi_open}
\end{aligned}
\end{equation}
where
\begin{equation}
    \phi_\text{fp} = \delta_1\frac{T_1^\pi}{2} + \delta_R\left(T_1^\pi  + T_s + \frac{1}{\Omega_R}\right)
\end{equation}
is the phase contribution due to having finite pulse times and Rabi frequencies. We drop a $\pi$ phase offset, as it does not impact the analysis here and is set by our arbitrary choice of initial rotation axis.
The first term in Eq.~\eqref{eq:phi_open} is a phase accrual during the dark time $\tau$ that is dependent on the laser frequency difference from $\omega_0$.
Notably, in the DF configuration, this term is insensitive to atomic velocity.
The effect of the finite excitation pulse durations and the temporal separation between pulses is captured in $\phi_\text{fp}$ and results in a detuning-dependent phase offset that leads to phase smearing but not decoherence.
The opening phase $\phi_{ge}^\text{open}(\tau)$ is equal to $\varphi(\tau)$ from Eq.~\eqref{eq:seq_open} of the main text.

\subsubsection{Output phase of a Ramsey sensor}

We now extend the opening sequence to a complete Ramsey sensor by adding a closing sequential excitation. The closing excitation consists of a Raman $\pi$-pulse, an inter-pulse free-evolution with duration $T_s$, and a single-photon $\pi/2$-pulse. 
During the dark time $\tau$, the $\ket g-\ket{s_+}$ coherence picks up a phase $\phi_{1}$ from laser 1 to trace out the Ramsey fringe. The total propagator is
\begin{align}
    U_\text{Ramsey}(\tau) = U_1^{\pi/2}\, U_f(T_s)\, U_R^\pi\, U_\text{open}(\tau),
\end{align}
where we absorb $\phi_1$ into the free evolution between pulses as $c_g \to c_g e^{-i\phi_1}$ during the propagation under $U_f(T_s)$. 
The closing $\pi/2$-pulse maps the accumulated $\ket g-\ket e$ relative phase into a population imbalance, with the ground state population at readout taking the form
\begin{equation}
    P_g(\Phi) = \frac{1}{2}\Big(1-C\cos(\Phi)\Big) ,
\end{equation}
where $C$ is the Ramsey contrast and $\Phi$ is the net accumulated phase.

Readout of the relative ground state population at the end of the sequence provides a measurement of $|c_g(\Phi)|^2$ where
\begin{align}
    c_g(\Phi) = \bra{g}U_\text{Ramsey}(\tau)\ket{e}.
\end{align}
The Ramsey output phase can be written with three structurally distinct contributions,
\begin{equation}\label{eq:Phi_ramsey}
    \Phi = (\delta_1+\delta_R)\tau + (\phi_A + \phi_B) + \phi_1,
\end{equation}
with
\begin{equation}
\begin{aligned}
    \phi_A + \phi_B =&\, \delta_1\frac{T_1^\pi}{2} + \delta_R\left(T_1^\pi  + T_s + \frac{1}{\Omega_R}\right) \\
    & + \delta_R\frac{T_R^\pi}{2} + \delta_1\left(T_R^\pi  + T_s + \frac{1}{\Omega_1}\right).
\end{aligned}
\end{equation}

Equation~\eqref{eq:Phi_ramsey} is the Ramsey output phase used in Sec.~\ref{sec:ramsey} of the main text. The first term corresponds to the sensor phase in Eq.~\eqref{eq:phi_sense} proportional to the laser frequency difference from $\omega_0$ and is the term in which Doppler dephasing manifests.
The second term is the finite-pulse correction term set by the pulse and inter-pulse durations that contributes a detuning-dependent but fixed phase offset.
The final term is the well-defined laser 1 phase shift that is used for tracing out Ramsey fringes.

\bibliography{ref}

\end{document}